\documentclass{ieeeaccess}
\usepackage{cite}
\usepackage{algorithmic}
\usepackage{graphicx}
\usepackage{textcomp}

\usepackage{amsmath}
\usepackage{amsfonts}
\usepackage{amssymb}
\usepackage{graphicx}
\usepackage[cmintegrals]{newtxmath}

\usepackage{subcaption}
\usepackage{bigstrut}
\usepackage{xcolor}
\usepackage{soul}

\usepackage{multirow}

\def\BibTeX{{\rm B\kern-.05em{\sc i\kern-.025em b}\kern-.08em
    T\kern-.1667em\lower.7ex\hbox{E}\kern-.125emX}}
\begin{document}
\history{Date of publication xxxx 00, 0000, date of current version xxxx 00, 0000.}
\doi{10.1109/ACCESS.2017.DOI}

\title{Site Diversity in Downlink Optical Satellite Networks Through Ground Station Selection}
\author{\uppercase{Eylem Erdogan}\authorrefmark{1}, \IEEEmembership{Senior Member, IEEE},
\uppercase{Ibrahim Altunbas}\authorrefmark{2},  \IEEEmembership{Senior Member, IEEE}, \uppercase{Gunes Karabulut Kurt}\authorrefmark{2,4},  \IEEEmembership{Senior Member, IEEE}, \uppercase{Michel~Bellemare}\authorrefmark{3},
\uppercase{Guillaume~Lamontagne}\authorrefmark{3}, \uppercase{Halim Yanikomeroglu}\authorrefmark{4} 
\IEEEmembership{Fellow, IEEE}}
\address[1]{Department of Electrical and Electronics Engineering, Istanbul Medeniyet University, 34700, Uskudar, Istanbul, Turkey  }
\address[2]{Electronics and Communication Engineering Department, Istanbul Technical University, 34469, Maslak, Istanbul, Turkey}
\address[3]{MDA, Satellite Systems, Sainte-Anne-de-Bellevue, H9X 3R2, QC, Canada}
\address[4]{Department of Systems and Computer Engineering, Carleton University, Ottawa, K1S 5B6, ON, Canada}
\tfootnote{This work was supported in part by the Optical Satellite Communications Consortium Canada (OSC).}

\markboth
{Author \headeretal: Preparation of Papers for IEEE TRANSACTIONS and JOURNALS}
{Author \headeretal: Preparation of Papers for IEEE TRANSACTIONS and JOURNALS}

\corresp{Corresponding author: Eylem Erdogan (e-mail: eylem.erdogan@medeniyet.edu.tr).}

\begin{abstract}
Recent advances have shown that satellite communication (SatCom) will be an important enabler for next generation terrestrial networks as it can provide numerous advantages, including global coverage, high speed connectivity, reliability, and instant deployment. An ideal alternative for radio frequency (RF) satellites is its free-space optical (FSO) counterpart. FSO or laser SatCom can mitigate the problems occurring in RF SatCom, while providing important advantages, including reduced mass, lower consumption, better throughput, and lower costs. Furthermore, laser SatCom is inherently resistant to jamming, interception, and interference. Owing to these benefits, this paper focuses on downlink laser SatCom, where the best ground station (GS) is selected among numerous candidates to provide reliable connectivity and site diversity. To quantify the performance of the proposed scheme, we derive closed-form outage probability and ergodic capacity expressions for two different practical GS deployment scenarios. Thereafter, asymptotic analysis is conducted to obtain the overall site diversity order, and aperture averaging is studied to illustrate the impact of aperture diameter on the overall performance. Furthermore, we investigate the site diversity order for a constellation of satellites that are communicating with the best GS by using opportunistic scheduling. Finally, important design guidelines that can be useful in the design of practical laser SatComs are outlined.

\end{abstract}


\titlepgskip=-15pt

\maketitle

\section{Introduction}
Satellite communication (SatCom) has become an important part of aerial networks in recent years due to its capabilities, which include flawless wireless connectivity, wide service coverage, and high-fidelity services for all the users around the world. An important feature of SatCom is to simultaneously transfer the signal rapidly around the Earth by providing distance-insensitive point-to-multipoint communications \cite{cheruku2010satellite}. So far, satellites have been used for television coverage, data communication, navigation, weather forecasts, climate and environmental monitoring, space science, and so on \cite{timothy2003satellite}. Satellites can be divided into three main categories depending on their altitudes and orbit types, low Earth orbit (LEO), medium Earth orbit (MEO) and geostationary Earth orbit (GEO). GEO satellites are {located at about $36000$ km above the Earth's equator to provide wide coverage, whereas LEO satellites}, which circle around the Earth at lower altitudes, have faster rotations, require less power, and are cheaper than GEO and MEO satellites \cite{kota2003broadband}. The technology behind SatCom is mainly based on radio-frequency (RF) systems, where $100$ MHz to $50$ GHz frequencies are used depending on the types and applications of satellites. In recent years, SatCom has emerged to provide high-speed, seamless broadband Internet connectivity around the globe as many different companies have started to launch constellations of satellites. For instance, SpaceX's Starlink recently began launching LEO satellites operating at high frequencies \cite{Starlink}. Similarly, OneWeb has launched about $104$ LEO satellites in $2020$, which operate in the $K_a$-$K_u$ band of the frequency spectrum \cite{OneWeb}. These recent developments suggest that RF-SatCom will become a key enabler in the integration of aerial and terrestrial networks in future wireless communication systems. In RF-SatCom, the most important drawbacks are cost, regulatory restrictions and limited available bandwidth, as it requires high data rates and broader bandwidths to connect anyone at anytime. Furthermore, RF-SatCom is prone to interference, jamming, and interception, which pose security risks, especially for military communications.

As a solution to these problems, free-space optical (FSO) or laser SatCom has attracted considerable interest both in recent academic and industry publications \cite{kaushal2016optical}. In laser SatCom, $20$ to $375$ THz spectrum\footnote{{In laser SatCom, only a small portion of the frequency spectrum can be used due to huge atmospheric losses.}} can be used to provide very high throughput in satellite-to-ground (downlink) and ground-to-satellite (uplink) communications \cite{itu-r1}. Laser SatCom can also provide significant advantages compared to its RF counterpart, including smaller antennas, reduced mass, lower consumption, better throughput, and lower costs. Furthermore, the narrow beam used in optical systems can provide secure communication, and it is immune to jamming, interception, interference. In addition, laser SatCom does not require a spectrum operating license for frequency use due to its inherent nature \cite{ince2012digital}. Due to these advantages, laser SatCom is expected to become a key enabler for future optical satellite systems, particularly for satellite-to-ground (downlink) communications, where line-of-sight (LOS) connectivity can be established perfectly \cite{giggenbach2008optical}. In downlink laser SatCom, the major adverse effects are atmospheric turbulence, atmospheric attenuation, and angle of arrival (AoA) fluctuations \cite{itu-r1}. The latter can be attenuated by using variable-focus lenses, which can adjust the beam size \cite{mai2019mitigation}. Also, aperture averaging, where the scintillation is spatially averaged over the aperture, can be used to reduce the adverse effects of atmospheric turbulence \cite{agarwal2019unified}. Finally, atmospheric attenuation due to adverse weather conditions can be resolved by achieving site diversity, in which the number of optical links can be enhanced by using multiple ground stations \cite{fuchs2015ground, net2016approximation, lyras2018optimum, gong2019network}. It can also be resolved by attaining spatial diversity with the aid of multiple apertures on a single GS, using appropriate combining techniques, such as selection combining, maximum ratio combining or equal gain combining \cite{gopal2016performance,li2015performance}. 

Due to the advantages outlined above, optical satellites have become an important topic in the recent literature. In \cite{johari2017performance, viswanath2015analysis,alshaer2020performance, alshaer2019analysis}, the issue of optical ground-to-satellite (uplink) communication was considered, and important performance metrics, including outage probability, error probability, and link capacity, were obtained in the presence beam wandering, climatic effects, and atmospheric attenuation. By contrast, \cite{gopal2016performance} and \cite{le2019throughput} focused on downlink SatCom, where the former investigated the impact of spatial diversity and aperture averaging, and the latter examined the throughput. Furthermore, \cite{illi2020phy} and \cite{zedini2020performance} proposed using FSO communication as a feeder link where the ground station (GS) feeds the satellite through a high capacity link.

In this paper, we pursue a different line of inquiry by providing a detailed analysis for the downlink optical SatCom systems. More precisely, to deal with the adverse weather conditions, we try to maximize the site diversity by selecting the best GS that provides the best channel conditions. It is important to note that site diversity, which can be crucial for creating seamless connectivity between satellite-to-ground links, has already been discussed in the literature e.g., \cite{fuchs2015ground, net2016approximation, lyras2018optimum, gong2019network} and the references therein. However, these previous works focused on the optimum GS selection problem from the network layer point-of-view. To the best of our knowledge, none of these works have considered the physical layer performance of the optical SatCom with GS selection. Given the importance of site diversity, and to fill the gap in the literature, we focus on the physical layer performance of the downlink optical SatCom with multiple GSs\footnote{In the proposed setup, we assume that all GSs are connected through fibre-optic wires, and the GS with the highest communication reliability can share the information to the other GSs.}. To quantify the performance of the proposed setup, we consider an aggregate channel model consisting of atmospheric turbulence and atmospheric attenuation, where we obtained two performance indicators, outage probability and ergodic capacity. We also consider the scenario of a constellation of satellites in which the best satellite communicates with the best GS by using opportunistic scheduling to maximize the site diversity. More precisely, the paper makes the following specific contributions: 
\begin{itemize}
	\item We focus on the physical layer performance of the  downlink optical SatCom, where the best GS is selected among a set of $\mathcal{K}$ sites that are available for communication, to minimize the outage probability and to maximize the ergodic capacity. For this network, we consider an aggregate channel model consisting of turbulence induced fading, and atmospheric attenuation due to Mie scattering and geometrical scattering, and we obtain the instantaneous signal-to-noise ratio (SNR). 
	\item After obtaining the instantaneous SNR, we derive new closed-form outage probability and ergodic capacity expressions to characterize the overall performance of the proposed scheme. We further elaborate the system at high SNR to obtain the overall diversity order. 
	\item We provide two new GS deployment scenarios: ground level deployment and high ground windy weather deployment. We also provide some interesting system design guidelines and consider aperture averaging to mitigate the adverse effects of turbulence induced fading. 
	\item We also obtain the diversity order when a set of satellites form a constellation and communicate with the best GS by using opportunistic scheduling.
\end{itemize} 
The remainder of the paper is organized as follows: In Section II, the system model and problem formulation are outlined. In Section III, the outage probability and ergodic capacity analyses are presented, and the overall diversity order of the considered multi-site system is obtained. In Section IV, numerical results are provided, and Section V concludes the paper.  

\begin{figure*}[t!]
	\centering
	\includegraphics[width=6.5in]{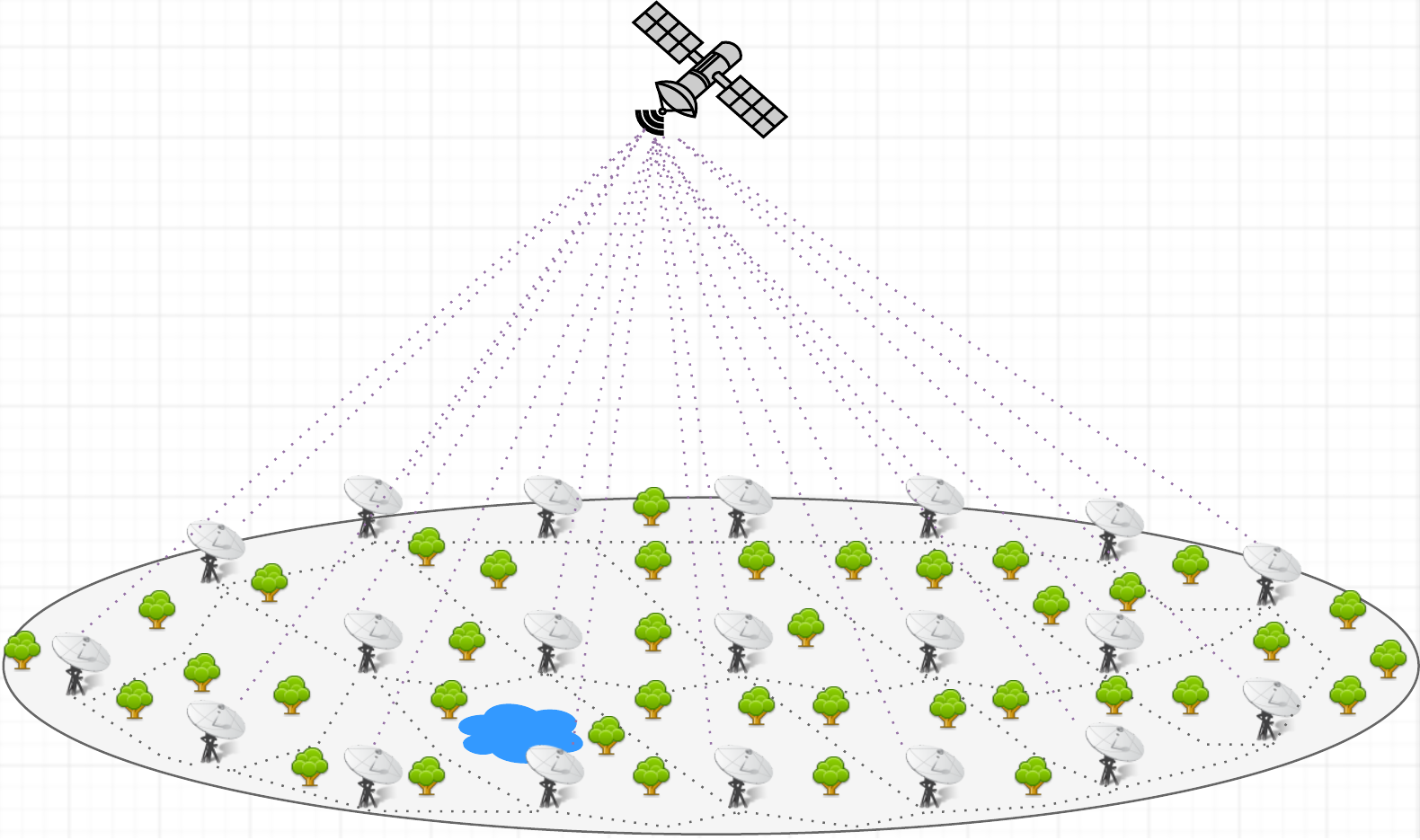}
	\caption{Optical downlink SatCom with multiple ground stations. }
	\label{fig_1}
\end{figure*}

\section{System Model and Problem Formulation}
We consider here a downlink optical SatCom where a low Earth orbit (LEO) satellite deployed on a circular orbit at $500$ km altitude seeks to communicate with the best GS selected among a set of $\mathcal{K}$ sites that are available for communication, as depicted in Fig. 1. {In the proposed model, we assume that all sites are in fixed locations providing perfect LOS connectivity, and the best GS with the highest signal-to-noise (SNR) ratio is selected to enhance the system performance by creating site diversity \cite{erdogan2019joint}. Prior to the data transmission, the satellite, which is moving at a speed of $8$ km/s, takes aim at the best GS, and the GS is aligned with the incoming beam to compensate for the bore-sight pointing errors. In this setup, the aggregated channel model is taken into consideration which consists of atmospheric attenuation ($I_j^{(a)}\big|_{j=1}^{\mathcal{K}}$) and turbulence induced fading ($I_j^{(t)}\big|_{j=1}^{\mathcal{K}}$). Mathematically speaking, the aggregated channel of the $j$-th GS can be expressed as \cite{awan2009cloud,dubey2020performance}
\begin{align}
I_j = I_j^{(a)}I_j^{(t)},  
\label{EQN:1}
\end{align}
and the instantaneous SNR ($\gamma_j$) can be expressed as
\begin{align}
\gamma_j = \frac{P_S}{N_0}I_j^2, 
\label{EQN:2}
\end{align}
where $P_S$ is the transmitted satellite signal power, and $N_0$ is the one sided noise power spectral density. The following subsections present the atmospheric attenuation and turbulence induced fading models for the proposed model. 
\subsection{Atmospheric Attenuation Model}
In downlink optical SatCom, two different scattering effects can be observed as the beam propagates to the GS. The first is Mie scattering, which redirects the transmitted signal from its intended direction when the signal wavelength is equal to the diameter of the particles in the medium. The second is geometrical scattering, which causes reflection, refraction, and scattering when the size of the particles in the medium is much greater than the signal wavelength\footnote{It is important to note that Rayleigh scattering may also adversely affect the optical SatCom. However, it can be negligible for systems operating below $375$ THz, as per the ITU-R report \cite{itu-r1}.}.  
\subsubsection{Atmospheric Attenuation Due to Mie Scattering}
Mie scattering is specified as the primary source of losses in downlink optical SatCom operating between $150$ and $375$ THz frequencies ($ \lambda = 0.8 - 2$ $\mu$m) and it is largely caused by microscopic particles of water \cite[Sec. (3.1)]{itu-r1}. The following expression, which can precisely model the Mie scattering effects, is appropriate for GSs located at altitudes between $0$ and $5$ km above the mean sea level \cite{itu-r1}: 
\begin{align}
\rho{'} = ah_E^3+bh_E^2+ch_E+d,
\label{EQN:3}
\end{align}
where $\rho{'}$ denotes the extinction ratio, $h_E$ stands for the height of the GS above the mean sea level (km), and $a$, $b$, $c$ and $d$ are the wavelength $\lambda$ ($\mu$m)-dependent empirical coefficients, which can be expressed as
\begin{align}
& a = -0.000545\lambda^2 + 0.002\lambda-0.0038 \nonumber \\
& b = 0.00628\lambda^2 -  0.0232\lambda + 0.0439 \nonumber \\
& c = -0.028\lambda^2 + 0.101\lambda  - 0.18 \nonumber \\
& d = -0.228\lambda^3 + 0.922\lambda^2 - 1.26\lambda + 0.719,
\label{EQN:4}
\end{align}
and the atmospheric attenuation due to Mie scattering $\Big(I_j^{(m)}\Big)$ can be expressed as\footnote{ Throughout the paper, we assume that the height of each GS above the mean sea level is the same. We also assume that all GSs are propagating at the same wavelength. However, the elevation angles and the propagation distances may vary depending on the location of each GS.}  
\begin{align}
I_j^{(m)} = \exp\Big(-\frac{\rho{'}}{\sin(\theta_j)}\Big),
\label{EQN:5}
\end{align}
where $\theta_j$ is the elevation angle of the $j$-th GS.

\begin{table}[t]
	\caption{Geometrical scattering parameters for various types of clouds at $1550$ nm. }
	\label{notations}
	\begin{center}
		\begin{tabular}{ |l| l| l| l|  } 
			\hline
			Cloud type & $N$ (cm$^{-3}$) & $\mathcal{L}_W$ ($g/m^{-3}$) & $V$ (km) \\
			\hline \hline
			Cumulus & $250$ & $1.0$ & $0.0280$ \\ 
			Stratus & $250$ & $0.29$ & $0.0626$ \\ 
			Stratocumulus & $250$ & $0.15$ & $0.0959$ \\ 
			Altostratus & $400$ & $0.41$ & $0.0369$ \\ 
			Nimbostratus & $200$ & $0.65$ & $0.0429$ \\ 
			Cirrus & $0.025$ & $0.06405$ & $64.66$ \\
			Thin cirrus & $0.5$ & $3.128\times 10^{-4}$ & $290.69$ \\
			\hline
		\end{tabular}
	\end{center}
	\label{Tab1}
\end{table}

\subsubsection{Atmospheric Attenuation due to Geometrical Scattering}
Geometrical scattering is used to model the attenuation that is close to the surface of the Earth and is caused by fog or dense clouds. In this model, visibility ($V_j$), which is an important factor for determining geometrical scattering, can be modeled in terms of liquid water content ($\mathcal{L}_{W_j}$) and cloud number concentration ($N_j$) as \cite{awan2009cloud}
\begin{align}
V_j = \frac{1.002}{(\mathcal{L}_{W_j}N_j)^{0.6473}}.
\label{EQN:6}
\end{align}
These parameters are summarized in Table \ref{Tab1} for various cloud formations. In geometrical scattering, the attenuation can be expressed by using the Beer-Lambert law,
\begin{align}
I_j^{(g)} = \exp(-\Theta_j L_j),
\label{EQN:7}
\end{align} 
where $L_j$ is the propagation distance, and $\Theta_j$ is the attenuation coefficient, which can be expressed as \cite[Sect. (3)]{ghassemlooy2019optical}
\begin{align}
\Theta_j = \Bigg(\frac{3.91}{V_j}\Bigg)\Bigg(\frac{\lambda}{550}\Bigg)^{-\psi_j},
\label{EQN:8}
\end{align}
where $\psi$ is the particle size related coefficient given according to Kim's model as
\begin{align}
\psi_j = 
\begin{cases}
1.6, & V_j>50\\
1.3  & 6<V_j<50 \\
0.16V_j+0.34 & 1<V_j<6\\
V-0.5 & 0.5<V_j<1\\   
0 & V_j<0.5,        
\end{cases}
\label{EQN:9}
\end{align}
and the atmospheric attenuation $I_j^{(a)}$ for the $j$-th GS can be expressed as \cite{johari2017performance}
\begin{align}
I_j^{(a)} = I_j^{(g)}I_j^{(m)} = \exp(-\sigma_j L_j) \exp\Big(-\frac{\rho'}{\sin(\theta_j)}\Big). 
\label{EQN:10}
\end{align}

\begin{table}[t]
	\caption{List of Notations and Parameters}
	\label{notations}
	\begin{center}
		\begin{tabular}{ |l|l|l| } 
			
			\hline
			Parameter & Definition \\
			\hline \hline
			
			$D_G$ & Hard receiver aperture diameter (m)   \bigstrut[t] \\
			
			$\mathcal{K}$ & Set of GSs that are available for communication \bigstrut[t] \\ 
			
			$L$ & Propagation distance  \bigstrut[t] \\ 
			
			$\zeta$ & Zenith angle  \bigstrut[t] \\ 
			
			$\theta$ & Elevation angle  \bigstrut[t] \\
			
			$v_r$ & Ground r.m.s. wind speed (m/s)  \bigstrut[t] \\
			
			$v_{g}$ & Ground wind speed (m/s)  \bigstrut[t] \\
			
			$h_E$ & Height of the GS above mean sea level (km) \bigstrut[t] \\
			
			$h$ & Altitude  \bigstrut[t] \\
			
			$H$ & Altitude of the satellite (m   \bigstrut[t] )\\
			
			$h_0$ & Height of the GS above ground level (m) \bigstrut[t] \\
			
			$\rho_c$ & Atmospheric correlation width \bigstrut[t] \\
			
			$\lambda$ & Wavelength \bigstrut[t] \\
			
			$\alpha, \beta$ & Shape parameters of the EW fading  \bigstrut[t] \\
			
			$\eta$  & Fading severity parameter of the EW fading  \bigstrut[t] \\
			
			$k$   &  Optical wave number \bigstrut[t] \\
			
			$V$ &  Visibility (km) \bigstrut[t] \\ 
			
			$\psi$ & Particle size related coefficient  \bigstrut[t] \\
			
			$N$ & Cloud number concentration \bigstrut[t] \\
			
			$\Theta$ & Attenuation coefficient \bigstrut[t] \\
			
			$\mathcal{L}_W$ & Liquid water content \bigstrut[t] \\
			
			$\sigma_R^2$  & Rytov variance \bigstrut[t] \\
			
			$C_n^2$ & Refractive index constant \bigstrut[t] \\
			
			$\sigma_I^2$  & Scintillation index \bigstrut[t] \\
			
			$\gamma_{th}$ &  Predefined threshold for acceptable communication quality \bigstrut[t] \\

			\hline
		\end{tabular}
	\end{center}
	\label{Tab2}
\end{table}
\normalsize

\subsection{Turbulence Induced Fading Model}
In this paper, the turbulence induced downlink channel experiences an exponentiated Weibull fading channel, where the corresponding probability density function (PDF) and cumulative distribution function (CDF) can be expressed as \cite{Barrios1}
\begin{align}
f_{I_j}^{(t)}(I) &= \frac{\alpha_j\beta_j}{\eta_j}\Bigg(\frac{I}{\eta_j}\Bigg)^{\beta_j-1}\exp\Bigg[-\Bigg(\frac{I}{\eta_j}\Bigg)^{\beta_j}\Bigg]\nonumber\\&\times \Bigg(1-\exp\Bigg[-\Bigg(\frac{I}{\eta_j}\Bigg)^{\beta_j}\Bigg]\Bigg)^{\alpha_j-1}
\label{EQN:11}	
\end{align}
and 
\begin{align}
{F}_{I_j}^{(t)}(I) = \Bigg(1-\exp\Bigg[-\Bigg(\frac{I}{\eta_j}\Bigg)^{\beta_j}\Bigg]\Bigg)^{\alpha_j}, 
\label{EQN:12}	
\end{align}
where $\alpha_j$, $\beta_j$ are the shape parameters and $\eta_j$ is the scale parameter of the $j$-th GS. The expressions for $\alpha_j$, $\beta_j$, and $\eta_j$ can be expressed as \cite{porras2013exponentiated}
\begin{align}
&\alpha_j = \frac{7.220\times \sigma_{I_j}^{2/3}}{\Gamma\Big(2.487\sigma_{I_j}^{2/6}-0.104\Big)}, \nonumber\\ & \beta_j = 1.012\Big(\alpha\sigma_{I_j}^{2}\Big)^{-13/25} + 0.142  \nonumber\\ & \eta_j = \frac{1}{\alpha\Gamma\Big(1 + 1/\beta_j\Big)g_1(\alpha_j,\beta_j)}, 	
\label{EQN:13}
\end{align}
where $g_1(\alpha_j,\beta_j)$ is the $\alpha$ and $\beta$ dependent constant variable, which can be written as \cite{porras2013exponentiated}
\begin{align}
g_1(\alpha_j,\beta_j)= \sum_{k=0}^{\infty}\frac{(-1)^k\Gamma(\alpha_j)}{k!(k+1)^{1+1/\beta_j}\Gamma(\alpha_j-k)}, 
\label{EQN:14}
\end{align}
and $\sigma_{I_j}^{2}$ denotes the scintillation index of the $j$-th GS, which can be given by \cite[Sect. (12)]{andrews2005laser}
\begin{align}
\sigma_{I_j}^2 = \exp\Bigg[\frac{0.49\sigma_R^2}{\big(1+1.11\sigma_{R_j}^{12/5}\big)^{7/6}}+\frac{0.51\sigma_{R_j}^2}{\big(1+0.69\sigma_{R_j}^{12/5}\big)^{5/6}}\Bigg] - 1, 
\label{EQN:15}
\end{align}
and the Rytov variance $\sigma_{R_j}^2$ can be expressed as \cite[Sect. (12)]{andrews2005laser}
\begin{align}
\sigma_{R_j}^2 = 2.25k^{7/6}\sec^{11/6}(\zeta_j)\int_{h_0}^H C_{n_j}^2(h)(h-h_0)^{5/6}dh,
\label{EQN:16}
\end{align}
where $k = \frac{2\pi}{\lambda}$ is the optical wave number, $\zeta_j$ is the zenith angle of the $j$-th GS, $h_0$ stands for the height of the GS above ground level, $H$ is the altitude of the satellite, and $C_{n_j}^2(h)$ is the altitude ($h$) dependent refractive index constant, which can be written as \cite{itu-r2}
\begin{align}
C_{n_j}^2(h) & = 8.148\times10^{-56}v_{r_j}^2h^{10}e^{-h/1000}+2.7\times10^{-16}e^{-h/1500} \nonumber\\ & + C_0e^{-h/100} \hspace{0.3cm} m^{-{2}/{3}},
\label{EQN:17}
\end{align}
where $v_{r_j} = \sqrt{v_{g_j}^2+30.69v_{g_j}+348.91}$ is the r.m.s ground wind speed in $m/s$ , $v_{g_j}$ is the ground wind speed in $m/s$ for the $j$-th GS, and $C_0 = 1.7\times 10^{-14}$ is the nominal value of the refractive index constant at ground level. Table \ref{Tab2} summarizes all notations and parameters.
\subsection{Problem Formulation}
This section formulates the GS selection strategy for the proposed setup. In what follows, the GS with the highest instantaneous SNR is selected to maximize the site diversity. Mathematically speaking, it can be formulated as 
\begin{align}
j^* = \arg\max_{1\leq j\leq \mathcal{K}}\Big[\gamma_j \Big],
\label{EQN:18}
\end{align}
where $j^*$ is the selected GS index. By doing so, outage probability ($P_{\text{out}}$) can be minimized as
\begin{align}
P_{\text{out}} = \Pr[\gamma \leq \gamma_{th}] = \Pr\bigg[\max_{1\leq j\leq\mathcal{K}}(\gamma_j) \leq \gamma_{th}\bigg],
\label{EQN:19}
\end{align}
where $\gamma_{th}$ is the predefined threshold for acceptable communication quality. Furthermore, ergodic capacity, which can assess the ergodic channel capacity, can be formulated as \cite{choi2018analysis}
\begin{align}
{\mathcal{C}}_{\text{erg}}  = {\mathbb{E}}\left[\log_2\Big(1+{\max_{1\leq j\leq\mathcal{K}}(\gamma_j)}\Big) \right] ,
\label{EQN:20}
\end{align}
where ${\mathbb{E}}\left[\cdot\right] $ is the expectation operation. It is important to note that it is almost impossible to find an exact ergodic capacity expression for the proposed scenario by using \eqref{EQN:20}. Therefore, with the aid of Jensen's inequality, we propose two approximate ergodic capacity bounds as given here;
\begin{subequations}\label{eq:litdiff}
	\begin{align}
	{\mathcal{C}}_{\text{erg}}^{\text{B}_1}  \approx \max_{1\leq j\leq\mathcal{K}}{\mathbb{E}}\left[\log_2(1+{\gamma_{j}}) \right],
	\label{EQN:21a}
	\end{align}
	\begin{align}
	{\mathcal{C}}_{\text{erg}}^{\text{B}_2}  \approx \log_2\Big({\mathbb{E}}\left[1+\max_{1\leq j\leq\mathcal{K}}({\gamma_{j}}) \right]\Big),
	\label{EQN:21b}
	\end{align}
\end{subequations}

\section{Performance Analysis}
This section derives new closed-form outage probability and ergodic capacity expressions for the proposed system. 

\subsection{Outage Probability Analysis}
Outage probability (OP) can be defined as the probability that the SNR will fall below a predefined threshold, $\gamma_{th}$, for acceptable communication quality. By substituting (\ref{EQN:12}) into (\ref{EQN:19}), with the aid of (\ref{EQN:18}), the OP can be expressed as
\begin{align}
P_{\text{out}} =  \prod_{j=1}^{\mathcal{K}}\Bigg(1-\exp\Bigg[-\Bigg(\frac{\gamma_{th}}{\Big(\eta_jI_j^{(a)}\Big)^2\bar{\gamma}_j^{(t)}}\Bigg)^{\beta_j/2}\Bigg]\Bigg)^{\alpha_j},
\label{EQN:22}
\end{align}
where $\bar{\gamma}_j^{(t)} = \frac{P_S}{N_0}\mathbb{E}\Big[\Big(I_j^{(t)}\Big)^2\Big]$ is the average SNR. By applying the Binomial theorem, and after a few manipulations, a tractable OP expression can be found as
\begin{align}
P_{\text{out}}= \prod_{j=1}^{\mathcal{K}}\sum_{\rho=0}^{\infty}\binom{\alpha_j}{\rho}(-1)^\rho\exp\Bigg[-\rho\Bigg(\frac{\gamma_{th}}{\Big(I_j^{(a)}\eta_j\Big)^2\bar{\gamma}_j^{(t)}}\Bigg)^{\frac{\beta_j}{2}}\Bigg]. 
\label{EQN:23}
\end{align}
To gain further insights about the system behavior, the OP can be analyzed at high SNR. To do so, we first invoke the high SNR assumption of $\exp(-x/a)\approx 1-x/a$ into (\ref{EQN:22}) as  
\begin{align}
P_{\text{out}}^{\infty} = \prod_{j=1}^{\mathcal{K}}\Bigg[\Bigg(\frac{\gamma_{th}}{\Big(\eta_jI_j^{(a)}\Big)^2\bar{\gamma}_j^{(t)}}\Bigg)^{\alpha_j\beta_j/2}\Bigg]. 
\label{EQN:24}
\end{align}
Then, if we express $\bar{\gamma}_j^{(t)} = \kappa_j\bar{\gamma}$, where $\kappa_j\big|_{j=1}^{\mathcal{K}}$ is constant, after a few manipulations, the above expression can be written as
\begin{align}
P_{\text{out}}^{\infty} = \prod_{j=1}^{\mathcal{K}}\Bigg[\Bigg(\frac{1}{\Big(\eta_jI_j^{(a)}\Big)^2\kappa_j}\Bigg)^{\alpha_j\beta_j/2}\Bigg] \Bigg(\frac{\gamma_{th}}{\bar{\gamma}}\Bigg)^{\sum_{j=1}^{\mathcal{K}}\alpha_j\beta_j/2}. 
\label{EQN:25}
\end{align}
At high SNR, the OP can be expressed as \cite{wang2003simple}
\begin{align}
P_{\text{out}}^{\infty} = \mathcal{G}_c(\bar{\gamma})^{-\mathcal{G}_d},
	\label{EQN:25x}
\end{align}
where $\mathcal{G}_c$ determines the shift of the curve in SNR. The diversity order $\mathcal{G}_d$ can be defined as the slope of the OP curve. For the considered multi-site system, $\mathcal{G}_d$ can be obtained as 
\begin{align}
	\mathcal{G}_d={\sum_{j=1}^{\mathcal{K}}\alpha_j\beta_j/2}
	\label{EQN:26x}
\end{align}
If we consider a practical system setup in which a set of $\mathcal{Z}$ satellites form a constellation and communicate with the best GS, which provides the highest SNR, by using opportunistic scheduling in which the best satellite providing best channel characteristics scheduled to communicate with the selected GS depending on the weather conditions, turbulence induced fading and LOS characteristics. In that case, the diversity order of the multi-satellite system can be obtained as
\begin{align}
\mathcal{G}_d^\text{C}=\sum_{z=1}^{\mathcal{Z}} \sum_{j=1}^{\mathcal{K}}\alpha_j\beta_j/2 \approx \mathcal{Z}\mathcal{K}\alpha\beta/2.
	\label{EQN:27x}
\end{align}

{\it Proof:} Please see Appendix A.

Furthermore, in this section, we propose two different practical deployment scenarios and investigate the OP performance of the proposed setup when there are $\mathcal{K} = 20$ set of GSs that are available for communication.

%

\subsubsection{Case 1 - Ground level deployment scenario}
In the first setup, we assume that all GSs are deployed at the ground level ($h_0 = h_E = 0$ m) above the mean sea and ground level, affecting from the nominal ground r.m.s. wind speed, which is $v_g = 2.8$ m/s. Assuming that the average SNR is $\bar{\gamma} = 24$ dB, and $\lambda = 1550$ nm, the OPs that are required for acceptable communication quality ($\gamma_{th} = 7$ dB) can be obtained as given in Case 1, Table \ref{Tab3}.  

\begin{table}[t]
	\caption{Outage probabilities for ground level deployment and high ground windy weather deployment scenarios}
	\label{notations}
	\begin{center}
		\begin{tabular}{ |l| l| l|  l| l| } 
			\hline
			Scenario & $\zeta$ & $N$ & $L_W$ & Outage probability   \\
			\hline \hline
			\multirow{4}{*}{Case 1} & $40^{\circ}$ &   $0.5$ & $3.128\times10^{-4}$ & $0.8166$ \\
			& $30^{\circ}$ &   $0.5$ & $3.128\times10^{-4}$ & $0.1619$  \\
			& $15^{\circ}$ &    $0.5$ & $3.128\times10^{-4}$ & $5.583\times10^{-4}$ \\
			& $0^{\circ}$ &   $0.5$ & $3.128\times10^{-4}$ & $2.165\times10^{-5}$  \\ \hline
			\multirow{4}{*}{Case 2}
			& $40^{\circ}$ &  $0.5$ & $3.128\times10^{-4}$ & 0.0986\\
			& $30^{\circ}$ &  $0.5$ & $3.128\times10^{-4}$ & $2.314\times10^{-4}$ \\
			& $15^{\circ}$ &   $0.5$ & $3.128\times10^{-4}$ & $2.27\times10^{-9}$\\
			& $0^{\circ}$ &   $0.5$ & $3.128\times10^{-4}$ & $1.28\times10^{-11}$ \\
			\hline
		\end{tabular}
	\end{center}
	\label{Tab3}
\end{table}

\subsubsection{Case 2 - High ground windy weather deployment scenario}
In the second setup, all GSs are deployed at $h_0 = 1000$ m above from the ground level, and $h_E = 1200$ m above from the mean sea level, where the wind is blowing at a speed of $11.176$ m/s.} Assuming that the average SNR is $\bar{\gamma} = 24$ dB and $\lambda = 1550$ nm, the OPs that are required for  acceptable communication quality can be summarized in Case 2, Table 3. We believe that these deployment scenarios can be used to provide practical insights about the system architecture, so that a system designer can get a quick idea about the overall system performance.

\begin{figure}[t!]
	\centering
	\includegraphics[width=3.5in]{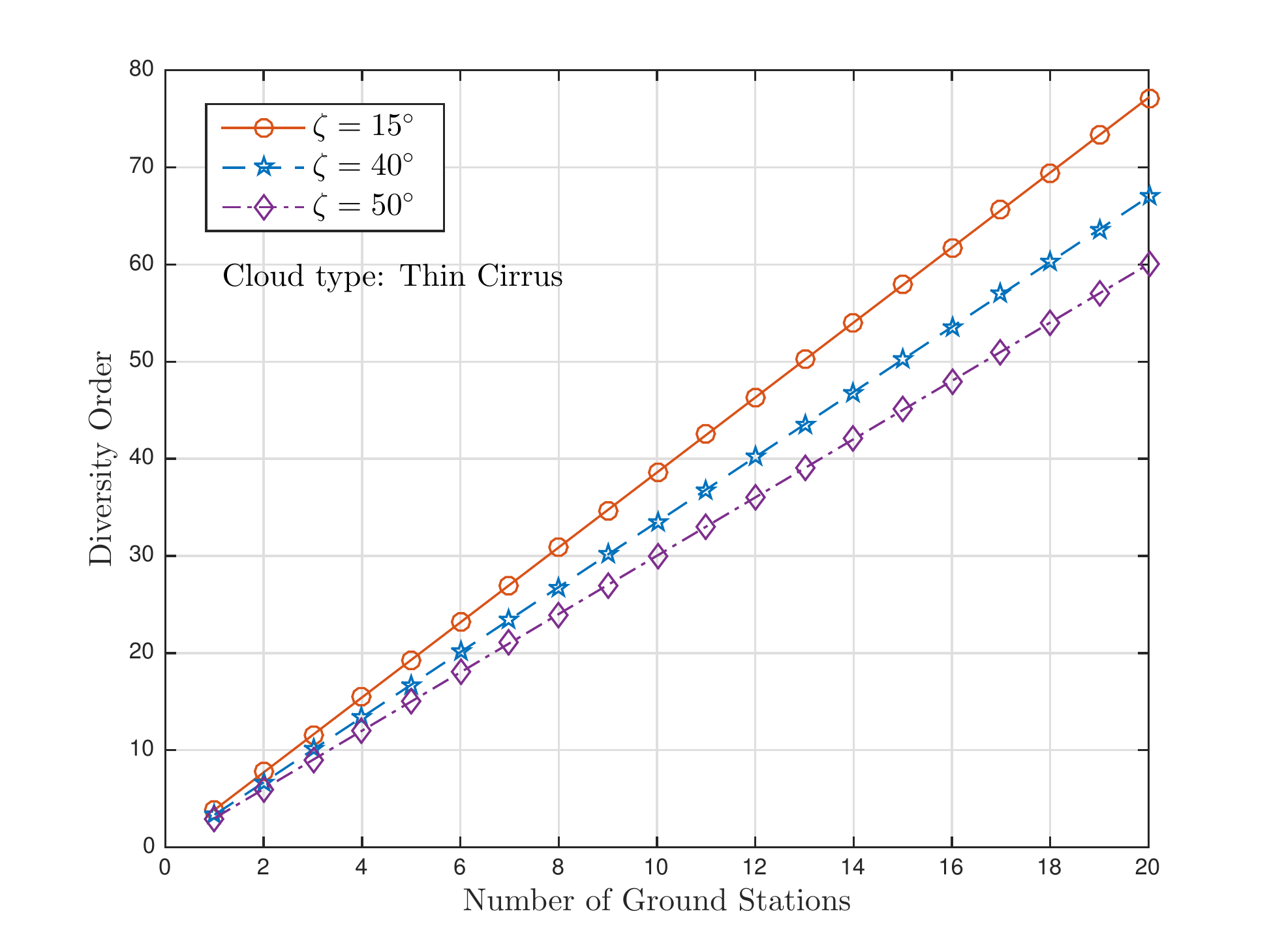}
	\caption{{Diversity order versus number of ground stations for the ground level deployment scenario}.}
	\label{fig_2}
\end{figure}

\begin{figure}
	\centering
	\includegraphics[width=3.5in]{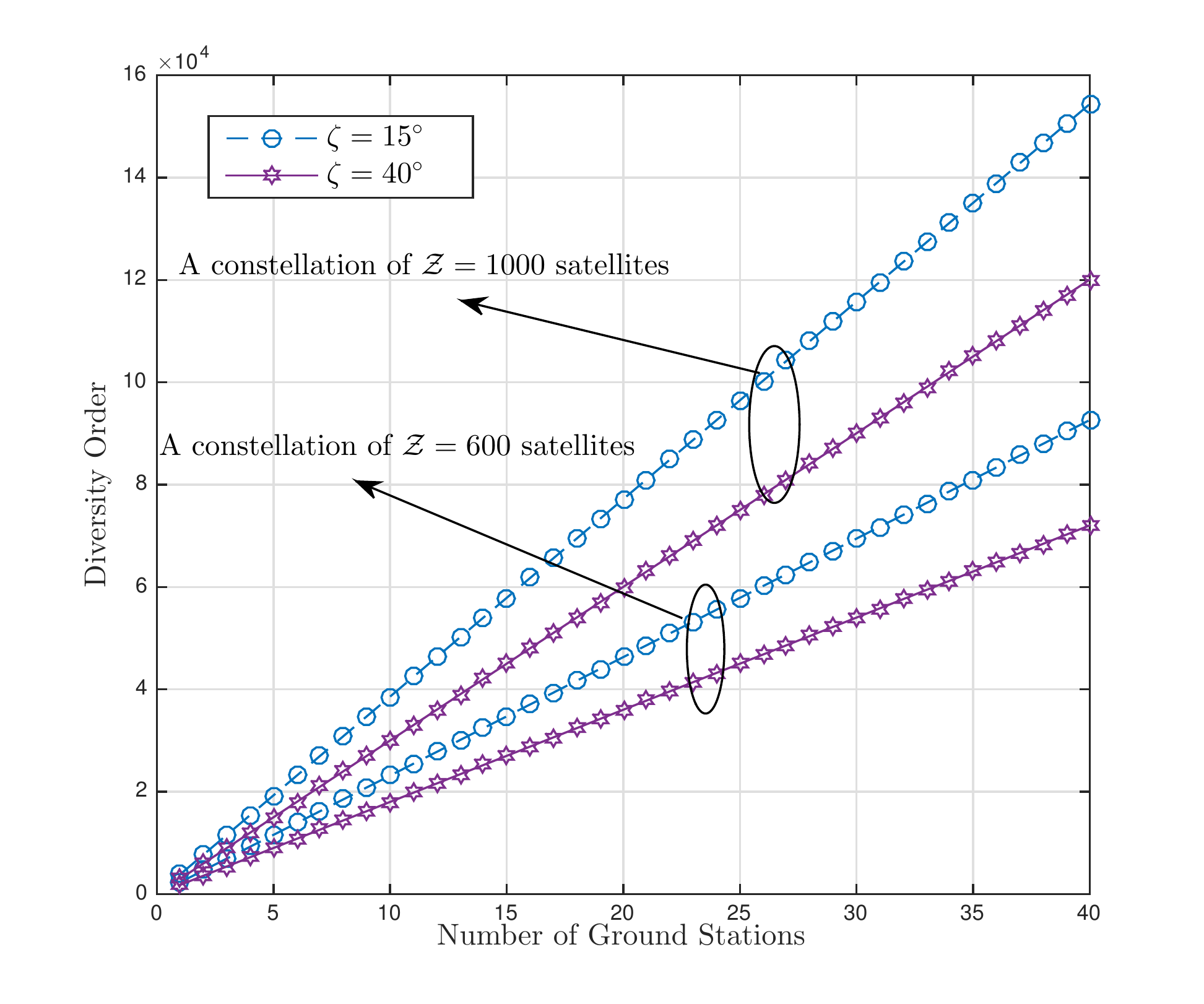}
	\caption{Diversity order versus number of ground stations for a constellation of satellites in the ground level deployment scenario.}
	\label{fig_2x}
\end{figure}

Table \ref{Tab3} shows the OP results for thin cirrus cloud formations, when the the satellite is on a $500$ km circular orbit. We can see from the table that, even though the wind speed increases in the high ground deployment, the overall OP performance of the proposed system enhances as both atmospheric turbulence and atmospheric attenuation effects reduces due to lower propagation distance. Furthermore, increasing the zenith angle shows that the GSs are affected by the attenuation and atmospheric turbulence at higher levels. Thereby, keeping the zenith angle small can boost the overall performance and can enhance the overall  diversity as can be observed from Fig. \ref{fig_2}. Furthermore, when $\mathcal{Z} = 600$ and $1000$ constellation of satellites are used, diversity order can be further improved as can be observed from Fig. \ref{fig_2x}. 



\vspace{0.2cm}
\subsection{Ergodic Capacity Analysis}
\subsubsection{First Approximate Bound on the Ergodic Capacity}
Ergodic capacity, expressed in bits/channel in use, can be defined with the aid of (\ref{EQN:21a}) as   
\begin{align}
{\mathcal{C}}_{\text{erg}}^{\text{B}_1}  = {\log_2(e)}\max_{1\leq j\leq\mathcal{K}}\Bigg\{\int_{0}^{\infty}\frac{1}{1+\gamma}\bar{F}_{\gamma_{j}}(\gamma)d\gamma\Bigg\},
\label{EQN:26}
\end{align}
where $\bar{F}_{\gamma_{j}}(\gamma) = 1-F_{\gamma_{j}}(\gamma)$ is the complementary CDF of $\gamma_{j}$, which can be given as
\begin{align}
\bar{F}_{\gamma_{j}}(\gamma) = \sum_{\rho=1}^{\infty}\binom{\alpha_j}{\rho}(-1)^{\rho+1}\exp\Bigg[-\rho\Bigg(\frac{\gamma}{\Big(\eta_jI_j^{(a)}\Big)^2\bar{\gamma}_j^{(t)}}\Bigg)^{\frac{\beta_j}{2}}\Bigg],
\label{EQN:27}
\end{align}
then, by invoking $\bar{F}_{\gamma_{j}}(\gamma)$ into (\ref{EQN:26}), ${\mathcal{C}}_{\text{erg}}$ can be expressed as 
\begin{align}
{\mathcal{C}}_{\text{erg}}^{\text{B}_1}  &= {\log_2(e)}\max_{1\leq j\leq\mathcal{K}}\Bigg\{\sum_{\rho=1}^{\infty}\binom{\alpha_j}{\rho}(-1)^{\rho+1}\int_{0}^{\infty}\frac{1}{1+\gamma}\nonumber\\&\times\exp\Bigg[-\rho\Bigg(\frac{\gamma}{\Big(\eta_jI_j^{(a)}\Big)^2\bar{\gamma}_{j}^{(t)}}\Bigg)^{\frac{\beta_j}{2}}\Bigg]d\gamma\Bigg\}.
\label{EQN:28}
\end{align}
To find the closed-form solution of the above integral, we first use the identity of  $\frac{1}{1+\gamma} = G_{1,1}^{1,1} \bigg[ \gamma \hspace{0.1cm} \bigg| \begin{matrix} 0 \\  0 \end{matrix} \bigg]$, and we can express ${\mathcal{C}}_{\text{erg}}$ as
\begin{align}
{\mathcal{C}}_{\text{erg}}^{\text{B}_1}  &= {\log_2(e)}\max_{1\leq j\leq\mathcal{K}}\Bigg\{\sum_{\rho=1}^{\infty}\binom{\alpha_j}{\rho}(-1)^{\rho+1}\int_{0}^{\infty}G_{1,1}^{1,1} \bigg[ \gamma \hspace{0.1cm} \bigg| \begin{matrix} 0 \\  0 \end{matrix} \bigg]\nonumber\\&\times\exp\Bigg[-\rho\bigg(\frac{\gamma}{\Omega_j}\bigg)^{\frac{\beta_j}{2}}\Bigg]d\gamma\Bigg\},
\label{EQN:29}
\end{align}
where $\Omega_j = \Big(I_j^{(a)}\eta_j\Big)^2\bar{\gamma}_j^{(t)}$ and $G_{c,d}^{a,b} \bigg[ \cdot \hspace{0.1cm} \bigg| \begin{matrix} \cdot \\  \cdot \end{matrix} \bigg]$ denotes the Meijer-G function \cite[eqn. 07.34.02.0001.01]{Wolfram}. By changing variables in the above integration as $\chi = \rho\bigg(\frac{\gamma}{\Omega_j}\bigg)^{\frac{\beta_j}{2}}$ and by using the identity of $\exp(-x) = G_{0,1}^{1,0} \bigg[ x \hspace{0.1cm} \bigg| \begin{matrix} - \\  0 \end{matrix} \bigg]$, the ergodic capacity can be written as
\begin{align}
{\mathcal{C}}_{\text{erg}}^{\text{B}_1}  &= {\log_2(e)}\max_{1\leq j\leq\mathcal{K}}\Bigg\{\sum_{\rho=1}^{\infty}\binom{\alpha_j}{\rho}(-1)^{\rho+1}\rho^{\frac{\beta_j}{2}}\Bigg(\frac{2\Omega_j}{\beta_j}\Bigg)\nonumber\\&\times\int_{0}^{\infty}\chi^{\frac{2}{\beta_j}-1}G_{1,1}^{1,1} \bigg[ \Omega_j\Bigg(\frac{\chi}{\rho}\Bigg)^{\frac{2}{\beta_j}} \hspace{0.1cm} \bigg| \begin{matrix} 0 \\  0 \end{matrix} \bigg] G_{0,1}^{1,0} \bigg[ \chi \hspace{0.1cm} \bigg| \begin{matrix} - \\  0 \end{matrix} \bigg]d\chi\Bigg\},
\label{EQN:30}
\end{align}
and with the aid of \cite[eqn. 07.34.21.0012.01]{Wolfram}, the closed form solution of the above expression can be obtained as 
\begin{align}
{\mathcal{C}}_{\text{erg}}^{\text{B}_1}  &= {\log_2(e)}\max_{1\leq j\leq\mathcal{K}}\Bigg\{\sum_{\rho=1}^{\infty}\binom{\alpha_j}{\rho}(-1)^{\rho+1}\rho^{\frac{\beta_j}{2}}\Bigg(\frac{2\Omega_j}{\beta_j}\Bigg)\nonumber\\ &\times H_{2,1}^{1,2} \bigg[ \Omega_j\rho^{\frac{\beta_j}{2}}\hspace{0.1cm} \bigg| \begin{matrix} (0,1),(1-\frac{2}{\beta_j},\frac{2}{\beta_j}) \\  (0,1) \end{matrix} \bigg]\Bigg\}.
\label{EQN:31}
\end{align}  
where $H_{p, q}^{m,n} \bigg[ \cdot \hspace{0.1cm} \bigg| \begin{matrix} \cdot,\cdot \\ \cdot,\cdot \end{matrix} \bigg]$ denotes the Fox H-function \cite{mathai2009h}.
\subsubsection{Second Approximate Bound on the Ergodic Capacity}
Another tight bound on the ergodic capacity can be obtained by using (\ref{EQN:21b}). First, recall that 
\begin{align}
{\mathcal{C}}_{\text{erg}}^{\text{B}_2} \approx \log_2\Big(1+{\mathbb{E}}\Bigg[\underbrace{\max_{1\leq j\leq\mathcal{K}}({\gamma_{j}})}_{\gamma} \Bigg]\Big).
\label{EQN:32}
\end{align}
Thereafter, $\mathbb{E}\left[\gamma\right]$ can be expressed as 
\begin{align}
\mathbb{E}\left[\gamma\right] = \int_0^\infty(1-F_\gamma(\gamma))d\gamma,
\label{EQN:33}
\end{align}
where $F_\gamma(\gamma)$ can be obtained very similarly to \eqref{EQN:23} after changing $\gamma_{th}$ with $\gamma$, and it can be approximately expressed as  
\begin{align}
F_\gamma(\gamma) \approx \sum_{\rho=0}^{\infty}\binom{\mathcal{K}\alpha}{\rho}(-1)^\rho\exp\Bigg[-\rho\Bigg(\frac{\gamma_{th}}{\Big(I^{(a)}\eta\Big)^2\bar{\gamma}^{(t)}}\Bigg)^{\frac{\beta}{2}}\Bigg].
\label{EQN:34}
\end{align}
By substituting (\ref{EQN:34}) into (\ref{EQN:33}), and after few manipulations, $\mathbb{E}\left[\gamma\right]$ can be obtained as 
\begin{align}
\mathbb{E}\left[\gamma\right] =  \sum_{\rho=1}^{\infty}\binom{\mathcal{K}\alpha}{\rho}(-1)^{\rho+1}\rho^{-2/\beta}\bar{\gamma}\Big(I^{(a)}\eta\Big)^2\Gamma(1+2/\beta). 
\label{EQN:35}
\end{align}
Finally, substituting \eqref{EQN:35} into \eqref{EQN:32}, ${\mathcal{C}}_{\text{erg}}^{\text{B}_2}$ can be easily obtained. 

\subsection{Aperture Averaging}
In downlink optical SatCom, the impact of scintillation can be great enough to limit the performance of GS receivers. To compensate for the impact of scintillation, an important enabler is aperture averaging. In downlink communication, aperture averaging takes place especially when the atmospheric correlation width $\rho_{c_j}$, which describes the effective diameter of the $j$-th aperture, is lower than the aperture diameter of the $j$-th GS ($D_{G_j}$), i.e, $\rho_{c_j}<D_{G_j}$. In this case, scintillation is spatially averaged over the aperture to reduce the adverse effects of scintillation. For the proposed setup, $\rho_{c_j}$ can be calculated as \cite[Sect. (12)]{andrews2005laser}
\begin{align}
\rho_{c_j} \approx \sqrt{\frac{45\times 10^3\sec(\zeta_j)}{k}}, \hspace{0.2cm} \sigma_{R_j}^2<<1, \hspace{0.2cm} 0\leq\zeta_j<50,
\label{EQN:36}
\end{align}
where $\rho_{c_j}$ represents the diameter of the point-like aperture for the GS. For example, when $\zeta_j = 40^{\circ}$, $\rho_{c_j} \approx 1.204$ cm shows the point-like aperture size for the optical downlink SatCom operating at $\lambda = 1.55$ nm wavelength. In aperture averaging, the aperture diameter dependent scintillation index can be expressed as \cite[Sect. (12)]{andrews2005laser}
\small
\begin{align}
&\sigma_{I_j}^2 = 8.7k^{7/6}(H-h_0)^{5/6}\sec^{11/6}(\zeta_j) \nonumber \\ \times &\Re\Bigg\{ \int_{h_0}^H C_n^2(h)\Bigg[\Bigg(\frac{kD_G^2}{16L} + i\frac{h-h_0}{H-h_0}\Bigg)^{5/6} - \Bigg(\frac{kD_{G_j}^2}{16L_j}\Bigg)^{5/6}\Bigg]dh \Bigg\},
\label{EQN:37}
\end{align}
\normalsize
where $\Re\{\cdot\}$ represents the real-valued terms. For example, considering $\zeta_j = 40^{\circ}$, $H=5\times10^5$ m, and thin cirrus cloud formations for the ground level deployment scenario,  $D_g = 20$ cm aperture size can boost the overall outage performance from $ 6.884\times10^{-7}$ to $3.441\times10^{-9}$ at $30$ dB average SNR.

\begin{table}[t]
	\caption{List of Parameters and Values}
	\label{notations}
	\begin{center}
		\begin{tabular}{ |l|l|l| } 
			
			\hline
			Parameters & Values \\
			\hline \hline
			
			$\mathcal{K}$ & 10, 20 \bigstrut[t] \\ 
			
			$\zeta$ & $15^\circ$, $40^\circ$, $50^\circ$ \bigstrut[t] \\ 
			
			$\lambda$ & $1.55$ nm \bigstrut[t] \\
			
			$H$ & $5\times10^5$ m \bigstrut[t] \\
			
			$h_0$ & $0$ m, $1000$ m \bigstrut[t] \\
			
			$h_E$ & $0$ km, $1.2$ km  \bigstrut[t] \\
			
			$v_g$ & $2.8$ m/s, $11.176$ m/s  \bigstrut[t] \\
			
			$\gamma_{th}$ & $7$ dB \bigstrut[t] \\
			
			$C_0$  & $1.7\times 10^{-14}$ \bigstrut[t] \\

			\hline
		\end{tabular}
	\end{center}
	\label{Tab5}
\end{table}
\normalsize


%
\begin{figure}[t!]
	\centering
	\includegraphics[width=3.5in]{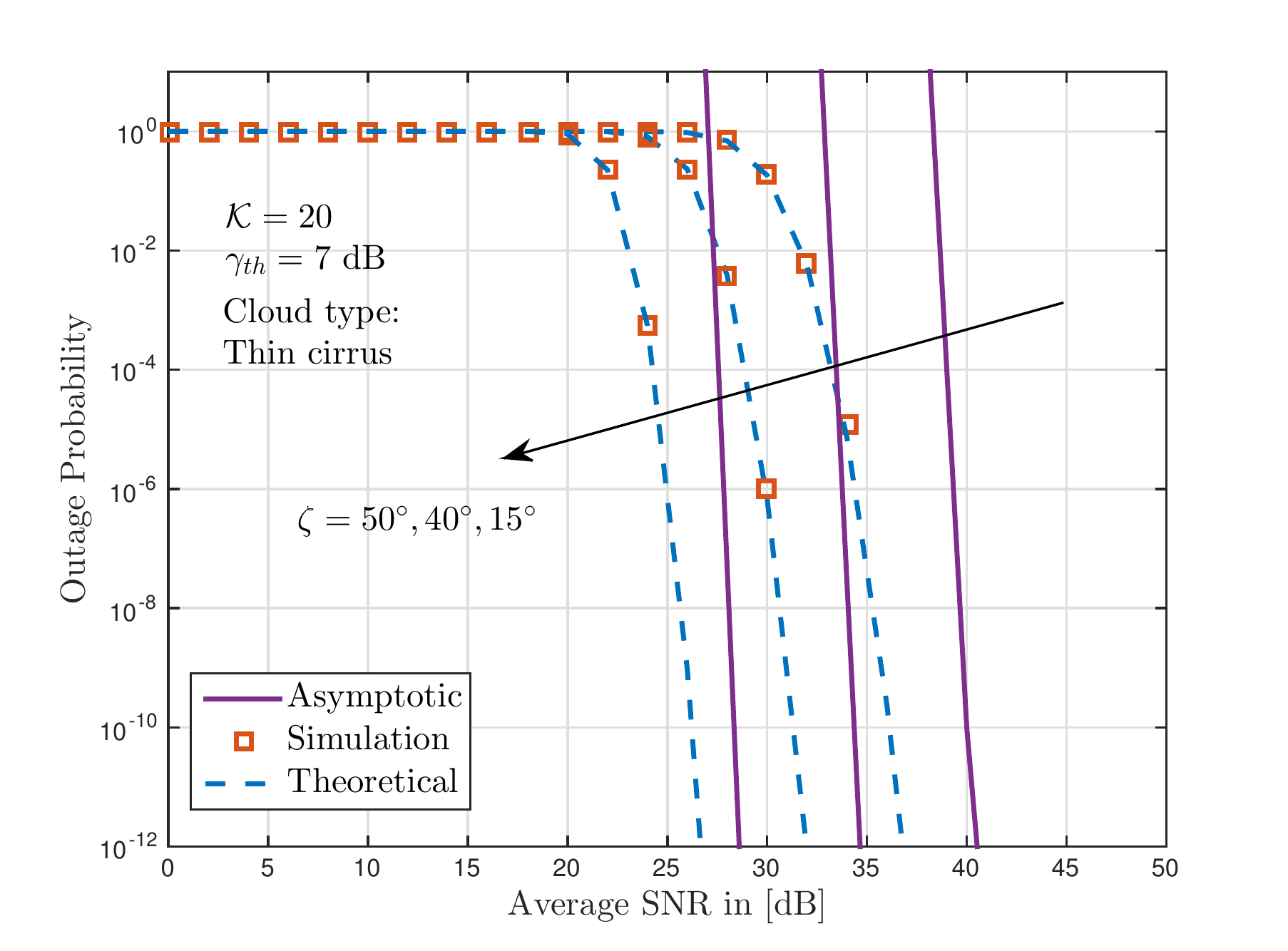}
	\caption{Outage probability performance of the proposed scheme for the ground level deployment scenario.}
	\label{fig_3}
\end{figure}
\begin{figure}[t!]
	\centering
	\includegraphics[width=3.5in]{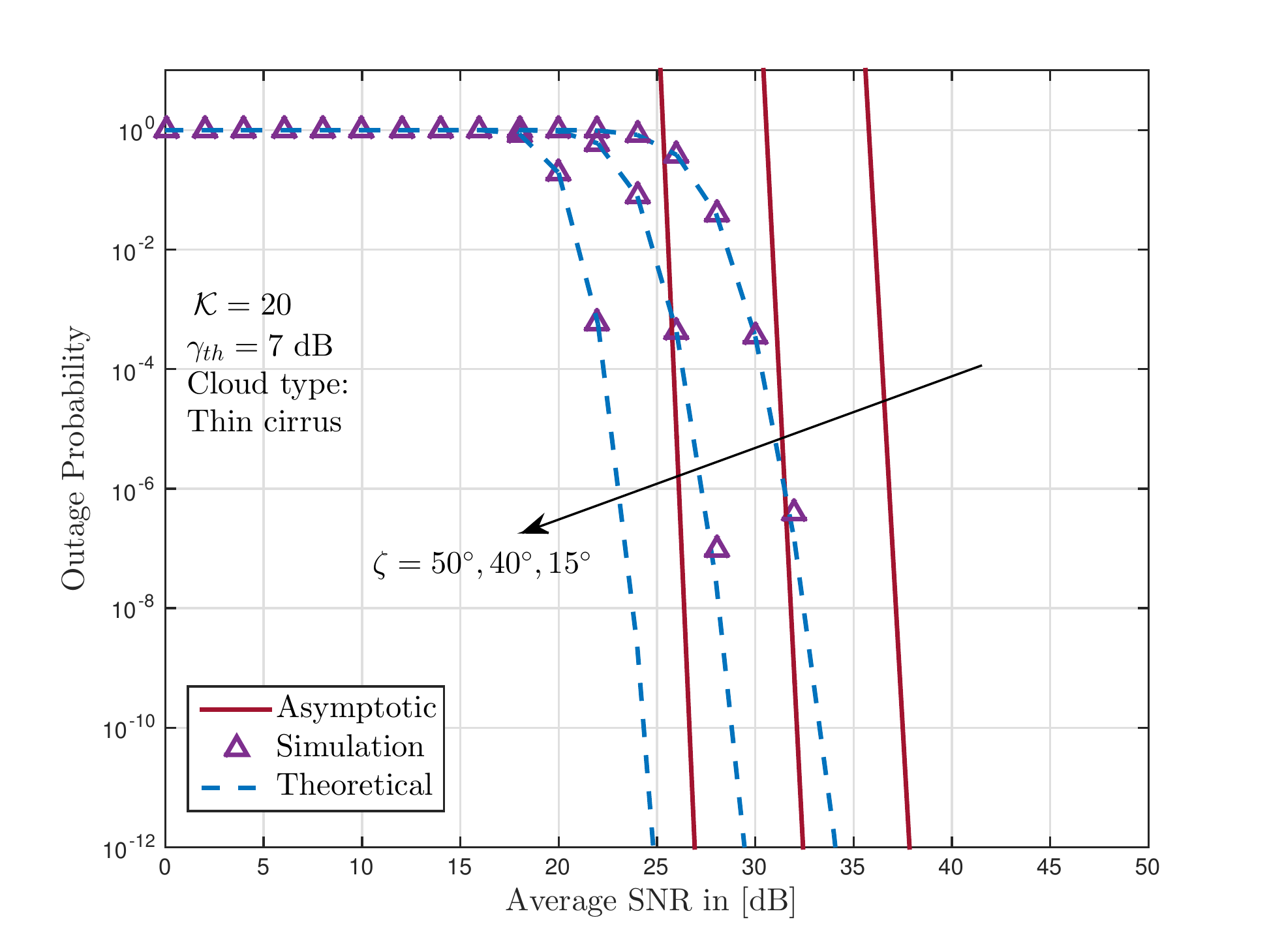}
	\caption{Outage probability performance of the proposed scheme for the high ground windy weather deployment scenario.}
	\label{fig_4}
\end{figure}

\section{Numerical Results} 
In this section, the theoretical results are first verified by a set of simulations. Then, the ground level deployment scenario is compared with the high ground windy weather deployment scenario in terms of outage probability and ergodic capacity. Finally, aperture averaging is illustrated in terms of outage probability, and important design guidelines are outlined for practical downlink laser SatComs. 

In all figures, the parameters are set as $\alpha_j = \alpha$, $\beta_j = \beta$, $\eta_j = \eta$, $I_j^{(a)} = I^{(a)}$, $I_j^{(t)} = I^{(t)}$  $\zeta_j = \zeta$, $\psi_j = \psi$, $C_{n_j}^2(h) = C_{n}^2(h)$, $\sigma_{I_j}^2 = \sigma_{I}^2$, $\sigma_{R_j}^2 = \sigma_{R}^2$ for notational brevity, without losing generality. Furthermore, in the ground level deployment scenario, the parameters are set as $h_0 = h_E = 0$ m, $v_g = 2.8$ m/s, whereas to demonstrate the high ground windy weather scenario, the parameters are set as $h_0 = 1000$ m,  $h_E = 1.2$ km, and $v_g = 11.176$ m/s. Finally, the SNR threshold is set to $\gamma_{th} = 7$ dB, and two different cloud forms, cirrus and thin cirrus, are used in the simulations together with three different $\zeta$ angles: $\zeta = 15^{\circ}, 40^{\circ}$ and $50^{\circ}$. All these parameters and values are illustrated in Table \ref{Tab5}.

\subsection{Verifications of the Theoretical Expressions}
Here we verify the theoretical results with the simulations. As we can see in Fig. \ref{fig_3} and \ref{fig_4}, the theoretical outage probability results, which are shown with dashed lines, are in good agreement with the marker symbols, which are generated by the simulations. Furthermore, both figures show that the overall outage performance of the proposed scheme can be enhanced remarkably by decreasing the $\zeta$ angle as expected. Finally, asymptotic curves, which are depicted with solid lines, show the overall diversity order, which is close to $\mathcal{G}_d=10\alpha\beta \approx 70$ at both figures.    

\begin{figure}[t!]
	\centering
	\includegraphics[width=3.5in]{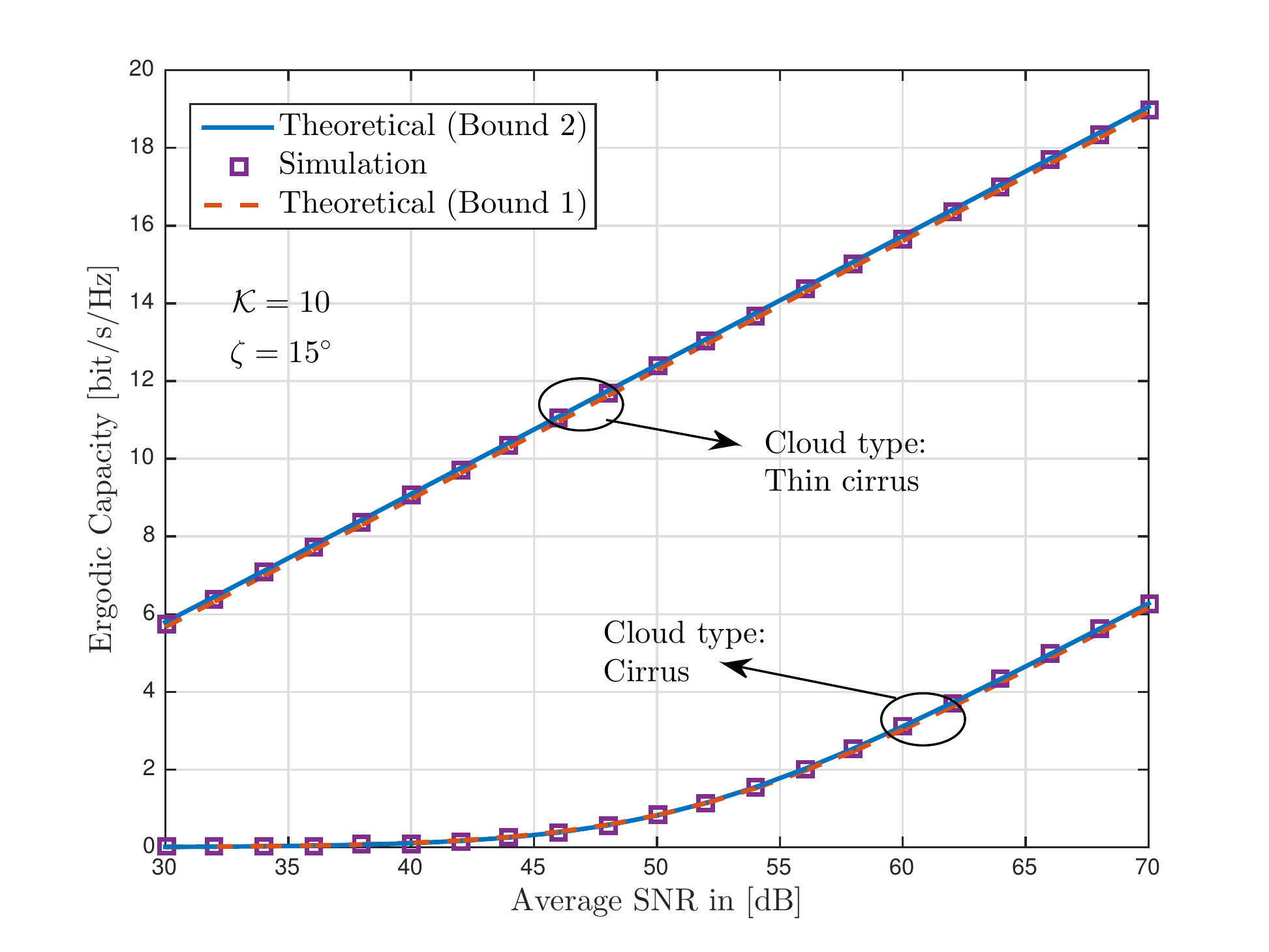}
	\caption{Ergodic capacity performance of the proposed scheme for the high ground windy weather deployment scenario.}
	\label{fig_5}
\end{figure}
\begin{figure}[t!]
	\centering
	\includegraphics[width=3.5in]{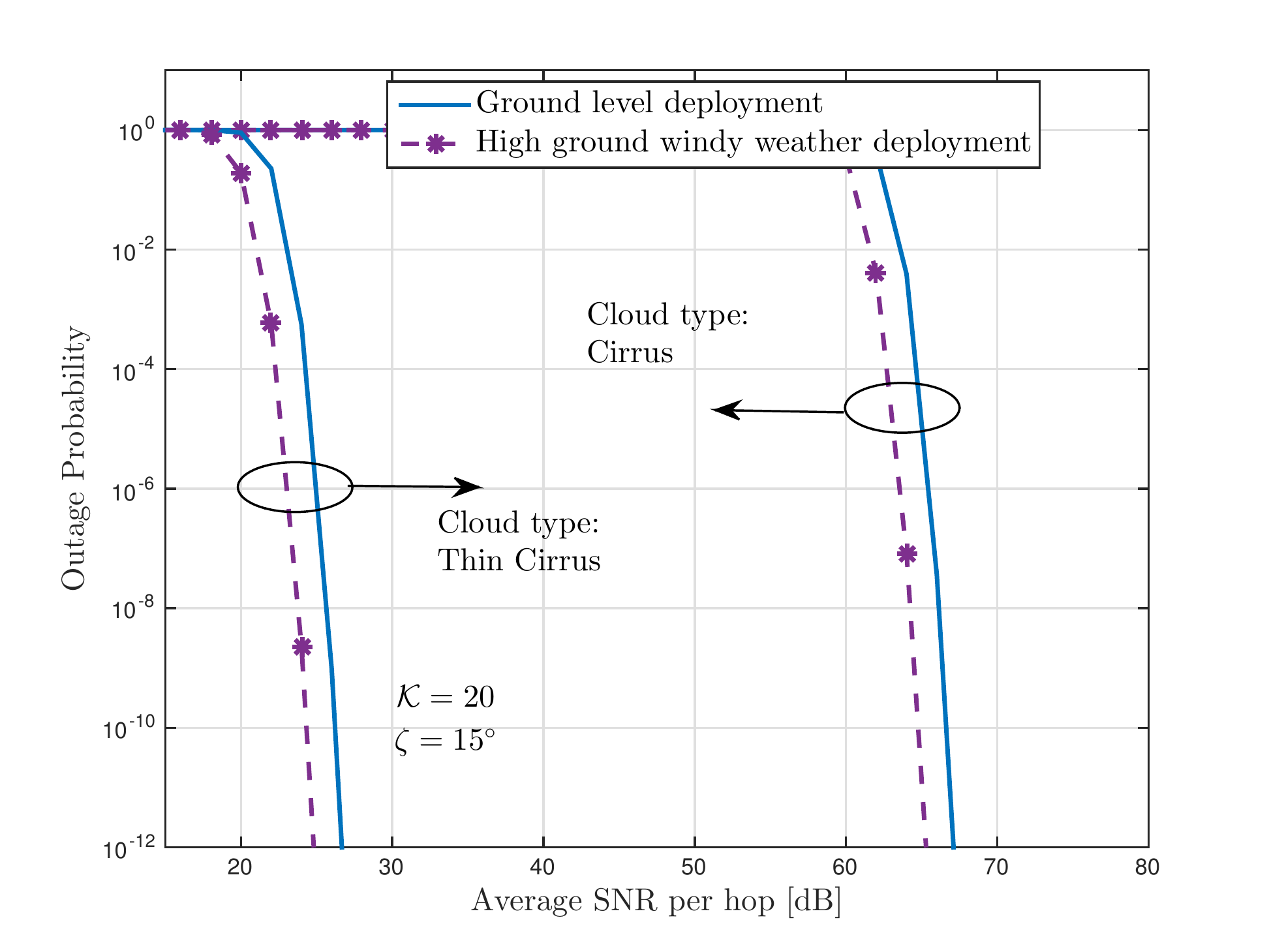}
	\caption{Comparison of ground level deployment with high ground windy weather deployment in terms of outage probability.}
	\label{fig_6}
\end{figure}
\begin{figure}[t!]
	\centering
	\includegraphics[width=3.5in]{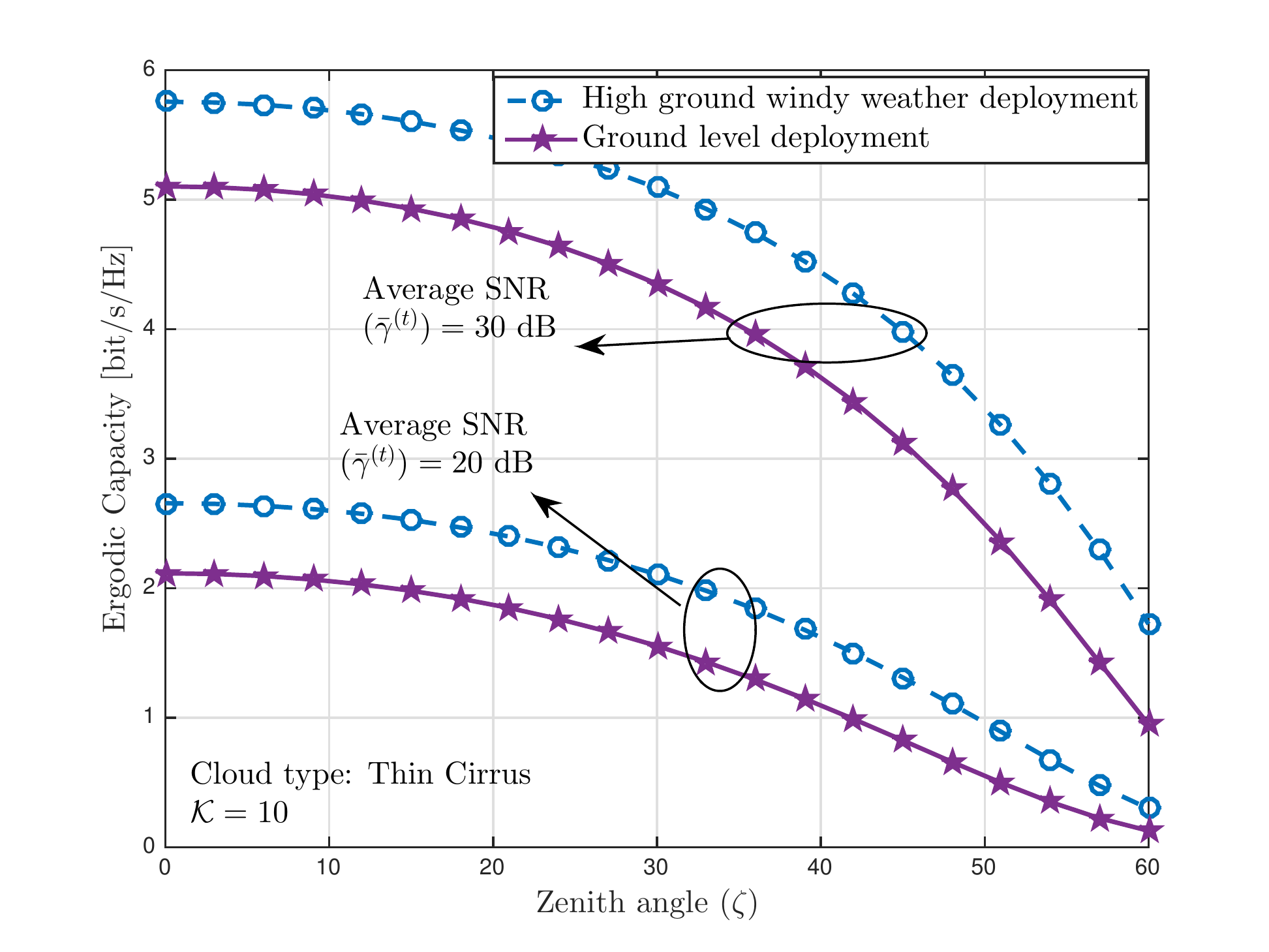}
	\caption{Comparison of ground level deployment with high ground windy weather deployment in terms of ergodic capacity.}
	\label{fig_7}
\end{figure}
\begin{figure}[t!]
	\centering
	\includegraphics[width=3.5in]{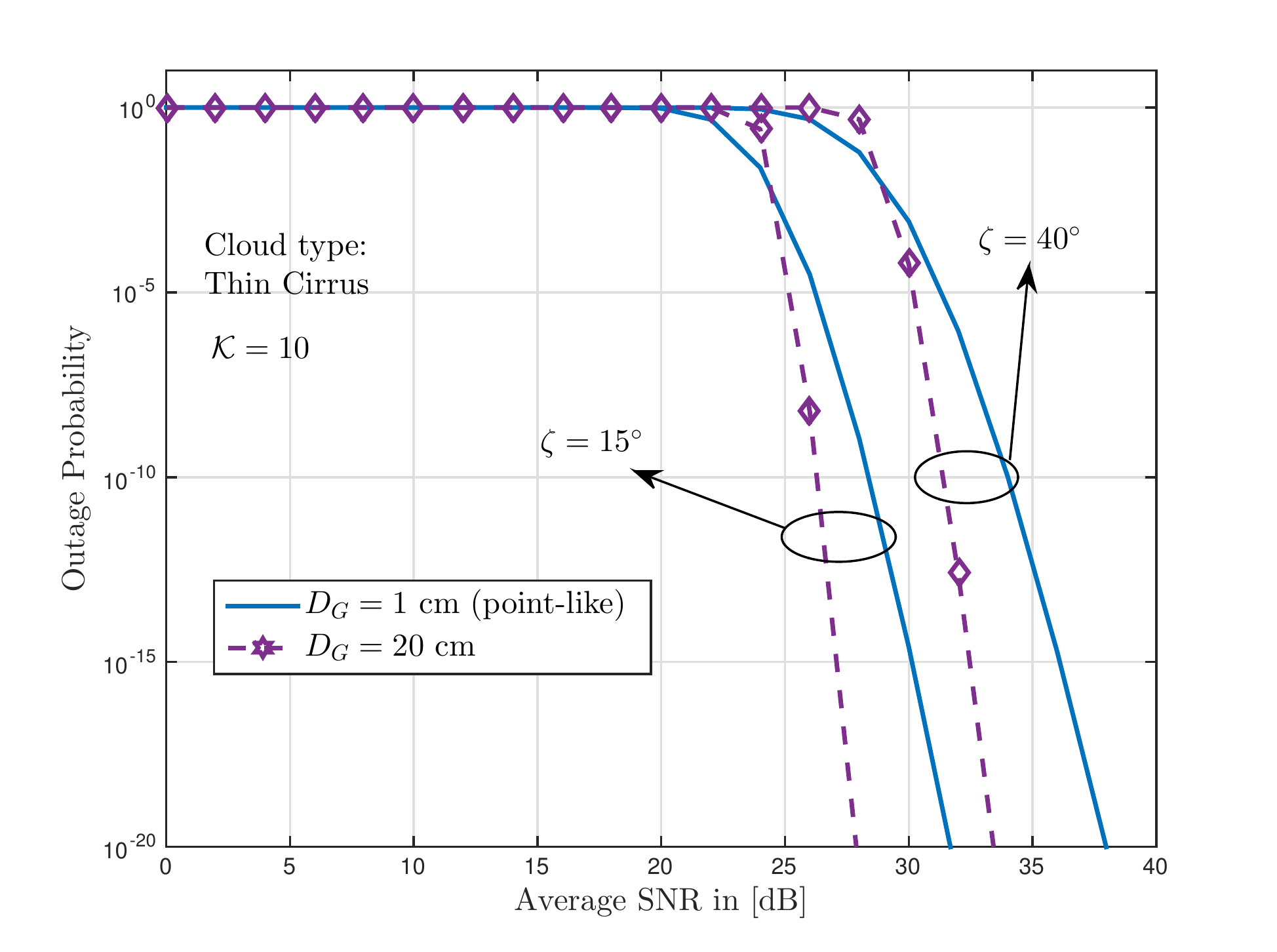}
	\caption{The impact of aperture averaging on the ground level deployment scenario in terms of outage probability.}
	\label{fig_8}
\end{figure}

Fig. \ref{fig_5}, on the other hand, illustrates the ergodic capacity performance of the proposed scheme for the high ground windy weather deployment scenario\footnote{Theoretical results that are evaluated by using bound 1 (shown with dashed lines) can be easily obtained by using Fox-H function in well-known software programs like MATHEMATICA or MATLAB. However, in a few cases, the Fox-H function may not work properly due to fractional fading severity values. For this reason, numerical integrations are used to verify the correctness of the first bound in certain cases.}. As we can see in the figure, theoretical bound 2, which is shown with solid lines, provides a tight upper bound, whereas theoretical bound 1, which are indicated with the dashed lines behaves like a tight lower bound with the marker symbols, which are the results of the simulations. Furthermore, thin cloud formations can bring up to $11$ bit/s/Hz capacity gain at $50$ dB average SNR as the atmospheric attenuation increases due to cirrus cloud formations, and results in losses on the ergodic capacity performance of the proposed scheme.

\subsection{Comparison of High Ground Windy Weather Deployment and Ground Level Deployment Scenarios}

Next, the ground level deployment scenario is compared with the high ground windy weather deployment scenario in terms of outage probability and ergodic capacity. Fig. \ref{fig_6} compares both schemes in terms of outage probability. As we can see in the figure, the high ground windy weather scenario outperforms the ground level counterpart both at thin cirrus and cirrus cloud formations, when $\zeta = 15^\circ$. Similarly, Fig. 8 shows the ergodic capacity performance of high ground windy weather deployment and ground level deployment scenarios in terms of zenith angle. As we can see from the figure, high ground windy weather deployment scenario outperforms its counterpart almost for all zenith angles. However, when $\zeta \rightarrow 60^\circ$, the gap between two deployment scenarios are closing and the ergodic capacity performance of the proposed scheme goes to $0$. Thereby, in downlink optical SatCom, almost no capacity gain can be obtained when $\zeta > 60^\circ$ as the optical signal is affected more from dense clouds, wind, and atmospheric turbulence induced fading.

\subsection{Impact of Aperture Averaging}
This section shows the impact of aperture averaging for the ground level deployment scenario in terms of outage probability. Fig. \ref{fig_8} shows that increasing the receiver aperture diameter up to $20$ cm increases the  diversity order and enhances the overall performance depending on the zenith angle and weather conditions. Also, interestingly, choosing the optimum zenith angle can yield a higher performance gain than aperture averaging.

\subsection{Constellation of Satellites}
This section illustrates the outage probability by considering a constellation of satellites that are communicating with the best GS by using opportunistic scheduling. In Fig. 10, outage probability is depicted for $\mathcal{\zeta} = 1000$ constellation of satellites by considering ground level deployment scenario. As can be observed from the figure, a constellation of satellites can bring up to $7$ dB average SNR gain at about $10^{-6}$ outage probability. However, the potential traffic congestion on the inter-satellite communication and the data delivery latency should be taken into consideration.

\begin{figure}[t!]
	\centering
	\includegraphics[width=3.5in]{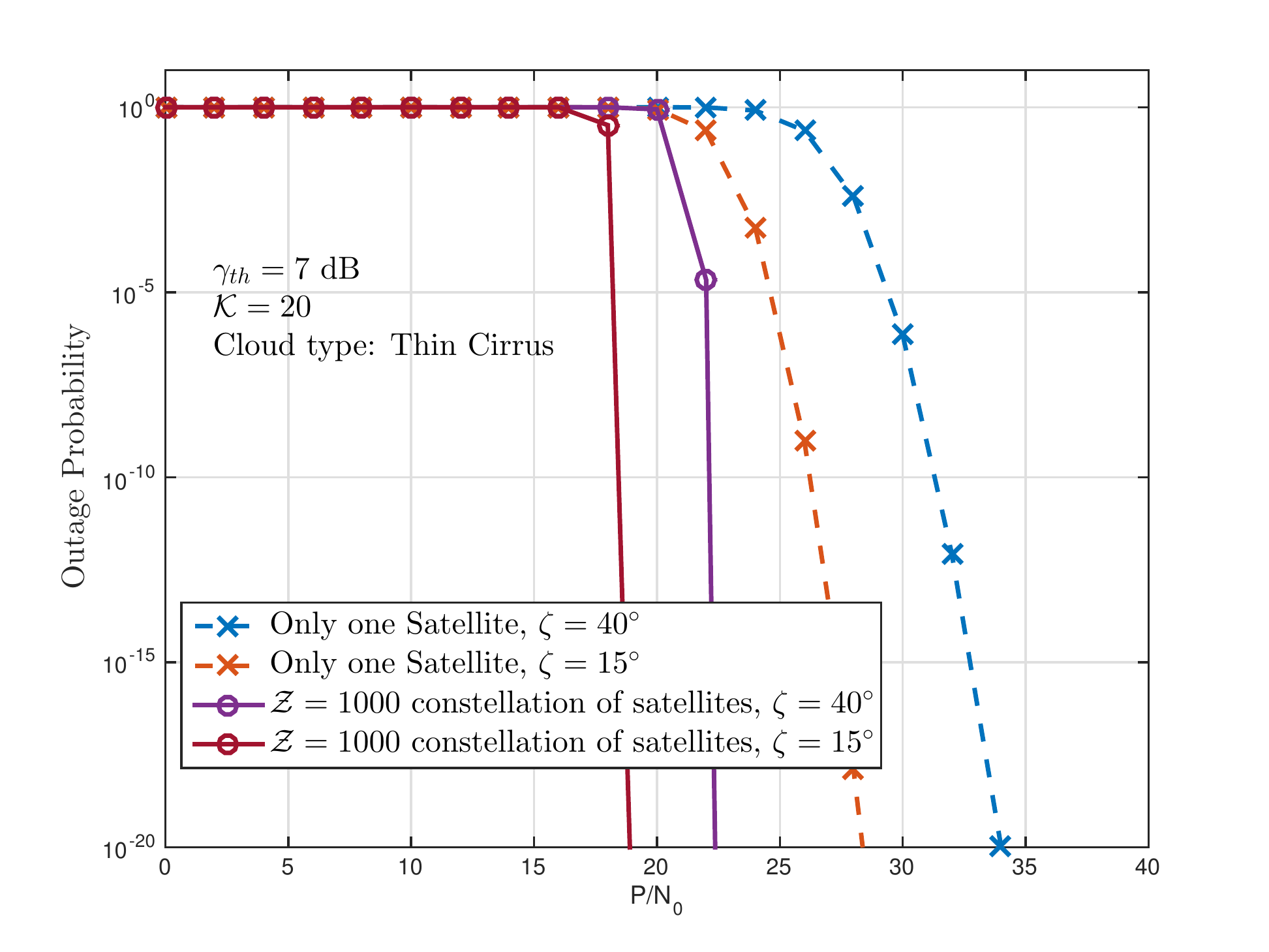}
	\caption{Outage probability performance of the proposed scheme for a $\mathcal{Z} = 1000$ constellation of satellites in the ground level deployment scenario.}
	\label{fig_10}
\end{figure}

\subsection{Design Guidelines}
Finally, we provide some important design guidelines that can be helpful in the design of downlink laser SatCom.

\begin{itemize}
	\item The simulations have shown that placing the GS to higher ground reduces the adverse atmospheric conditions and enhances the overall performance. So, the altitude of the GSs is of utmost importance in the design of downlink laser SatComs.
	\item The zenith angle has a direct impact on the design of downlink laser SatComs as the effect of atmospheric  turbulence is much lower when the zenith angle is a small value. For this reason, the zenith angle should be taken into consideration in the design of downlink laser SatCom.
	\item Aperture averaging can be an important enabler for enhancing the overall performance of the downlink laser SatComs, especially in the presence of adverse weather conditions, and when the zenith angle is higher than $\zeta = 30^{\circ}$.
	\item The simulations have shown that less than $10^{-10}$ outage probability can be achieved at $30$ dB when a set of $20$ GSs is used. This shows that creating site diversity through GS selection can be of utmost importance in downlink laser SatComs to enhance the overall performance.
	\item The simulations have shown that almost no capacity gain can be obtained when the zenith angle is greater than $60^\circ$ as the system is exposed to higher atmospheric turbulence induced fading and atmospheric attenuation due to increased propagation distance.   
	\item Forming a constellation of satellites can further enhance the performance gain up to $7$ dB at $10^{-5}$ outage probability and diversity order up to a thousand times when the constellations consists of $1000$ satellites. However, the potential traffic congestion on the inter-satellite communication and the data delivery latency should be taken into consideration in the design of satellite networks. 
\end{itemize}

\section{Conclusion}
This paper has focused on downlink laser SatComs, where the best GS is selected from among a set of $\mathcal{K}$ candidates to provide fully reliable connectivity and maximum site diversity. For the proposed structure, outage probability and ergodic capacity expressions were derived and asymptotic outage probability was conducted to provide the overall diversity order. Furthermore, we considered two different ground station deployment scenarios and investigated the impact of aperture averaging in terms of outage probability. Finally, important system design guidelines were provided to help in the design of downlink laser SatComs. The results have shown that the altitude of the GS and the zenith angle are of utmost importance in downlink laser SatCom.

\appendix
\section*{Appendix A: SITE DIVERSITY FOR CONSTELLATION OF SATELLITES}
	
Let us consider that a set of $\mathcal{Z}$ satellites form a constellation and communicating with the best GS which provides the highest signal-to-noise (SNR) with the aid of opportunistic scheduling. In that case, the best GS can be selected as 
\begin{align}
\{z^*,j^*\} = \arg {\max_{\substack{{1\leqslant j\leqslant \mathcal{K},}\\{1\leqslant z\leqslant \mathcal{Z}}}}}\Big[\gamma_{j,z} \Big],
\label{EQN:39}
\end{align}
where $j^*$ is the selected GS index and $z^*$ is the index of the selected satellite, and the outage probability can be minimized as
\begin{align}
P_{\text{out}} = \Pr[\gamma \leq \gamma_{th}] = \Pr\bigg[{\max_{\substack{{1\leqslant j\leqslant \mathcal{K},}\\{1\leqslant z\leqslant \mathcal{Z}}}}}(\gamma_{j,z}) \leq \gamma_{th}\bigg],
\label{EQN:40}
\end{align}
By substituting (\ref{EQN:12}) into (\ref{EQN:40}) with the aid of (\ref{EQN:39}), with few manipulations, the OP can be expressed as
\begin{align}
P_{\text{out}} &= \prod_{z=1}^{\mathcal{Z}}\prod_{j=1}^{\mathcal{K}}
\sum_{\rho=0}^{\infty}\binom{\alpha_{j,z}}{\rho}(-1)^\rho\nonumber\\ &\times\exp\Bigg[-\rho\Bigg(\frac{\gamma_{th}}{\Big(I_{j,z}^{(a)}\eta_{j,z}\Big)^2\bar{\gamma}_{j,z}^{(t)}}\Bigg)^{\frac{\beta_{j,z}}{2}}\Bigg], 
\label{EQN:41}
\end{align}
where $\bar{\gamma}_{j,z}^{(t)} = \frac{P_S}{N_0}\mathbb{E}\Big[\Big(I_{j,z}^{(t)}\Big)^2\Big]$ is the average SNR. To obtain the asymptotic behavior of the outage probability, we invoke the high SNR assumption of $\exp(-x/a)\approx 1-x/a$ into (\ref{EQN:41}) as
\begin{align}
P_{\text{out}}^{\infty} = \prod_{z=1}^{\mathcal{Z}}\prod_{j=1}^{\mathcal{K}}\Bigg[\Bigg(\frac{\gamma_{th}}{\Big(\eta_{j,z}I_{j,z}^{(a)}\Big)^2\bar{\gamma}_{j,z}^{(t)}}\Bigg)^{\alpha_{j,z}\beta_{j,z}/2}\Bigg],
\label{EQN:42}
\end{align}
and after several manipulations, the diversity order of the considered multi-site system can be obtained as given in (\ref{EQN:27x}). Hereto, the proof is completed.

\bibliographystyle{IEEEtran}
\bibliography{Refs.bib}

\begin{IEEEbiography}[{\includegraphics[width=1in,height=1.25in,clip,keepaspectratio]{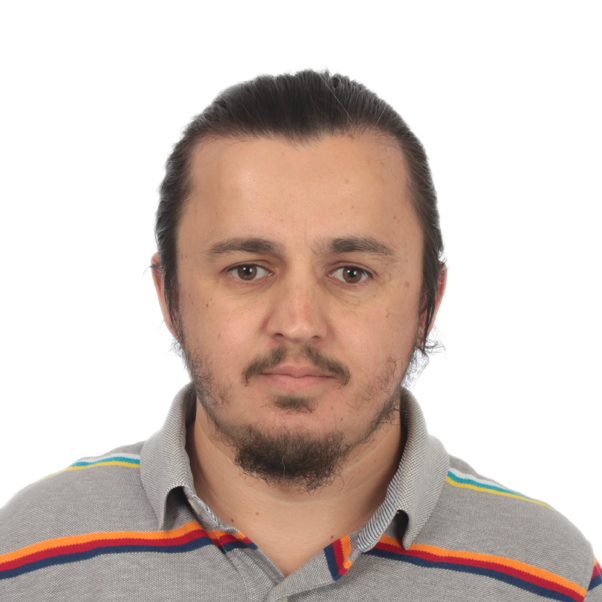}}]{Eylem Erdogan}  is currently an Associate Professor in the Department of Electrical and Electronics Engineering, Istanbul Medeniyet University. He was a Post-Doctoral Fellow in Electrical Engineering department, Lakehead University, Thunder Bay, ON, Canada from March 2015 to September 2016 and a visiting professor in Carleton University, Ottawa, Canada during summer 2019. His research interests are in the broad areas of wireless communications, the performance analysis of cooperative relaying in cognitive radio networks, unmanned aerial vehicle communications and networks, free space optical communications and optical satellite networks.

\end{IEEEbiography}

\begin{IEEEbiography}[{\includegraphics[width=1in,height=1.25in,clip,keepaspectratio]{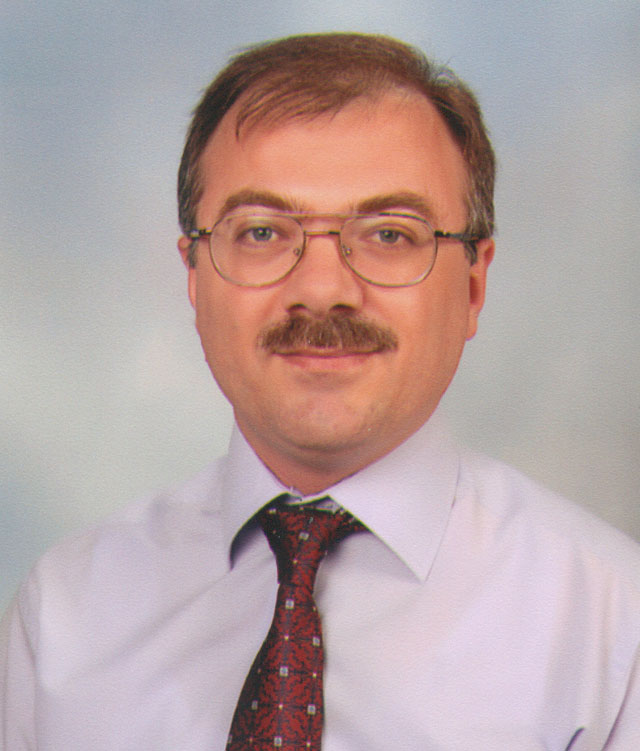}}]{Ibrahim Altunbas} is a Professor in the Electronics and Communication Engineering Department, Istanbul Technical University. He was a Visiting Researcher at Texas A\&M University (College Station, TX, USA) in 2001. He was a Postdoctoral Fellow and Visiting Professor at the University of Ottawa (ON, Canada) in 2002 and 2015 summer, respectively. His current research interests include spatial modulation, non-orthogonal multiple access, physical layer security, drone communications, and reconfigurable intelligence surface-based communication.
\end{IEEEbiography}

\begin{IEEEbiography}[{\includegraphics[width=1in,height=1.25in,clip,keepaspectratio]{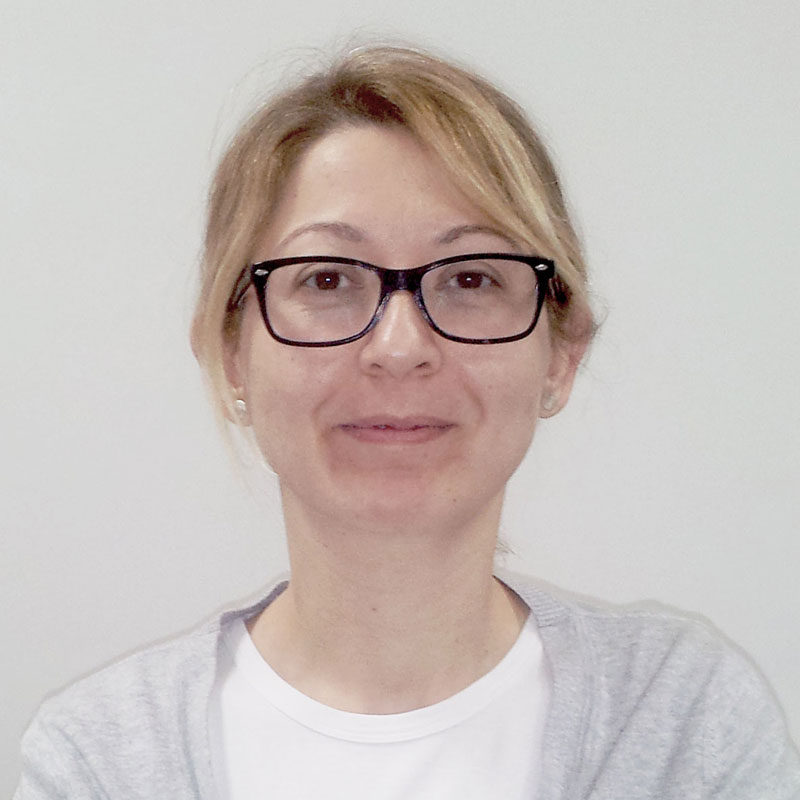}}]{Gunes Karabulut Kurt} received the B.S. degree with high honors in electronics and electrical engineering from the Bogazici University, Istanbul, Turkey, in 2000 and the M.A.Sc. and the Ph.D. degrees in electrical engineering from the University of Ottawa, ON, Canada, in 2002 and 2006, respectively. From 2000 to 2005, she was a Research Assistant with the CASP Group, University of Ottawa. Between 2005 and 2006, she was with TenXc Wireless, Canada. From 2006 to 2008, she was with Edgewater Computer Systems Inc., Canada. From 2008 to 2010, she was with Turkcell Research and Development Applied Research and Technology, Istanbul. Since 2010, she has been with Istanbul Technical University, where she currently works as a Professor. She is a Marie Curie Fellow. She is an adjunct research professor at Carleton University, also currently serving an Associate Technical Editor (ATE) of the IEEE Communications Magazine. 
\end{IEEEbiography}

\begin{IEEEbiography}[{\includegraphics[width=1in,height=1.25in,clip,keepaspectratio]{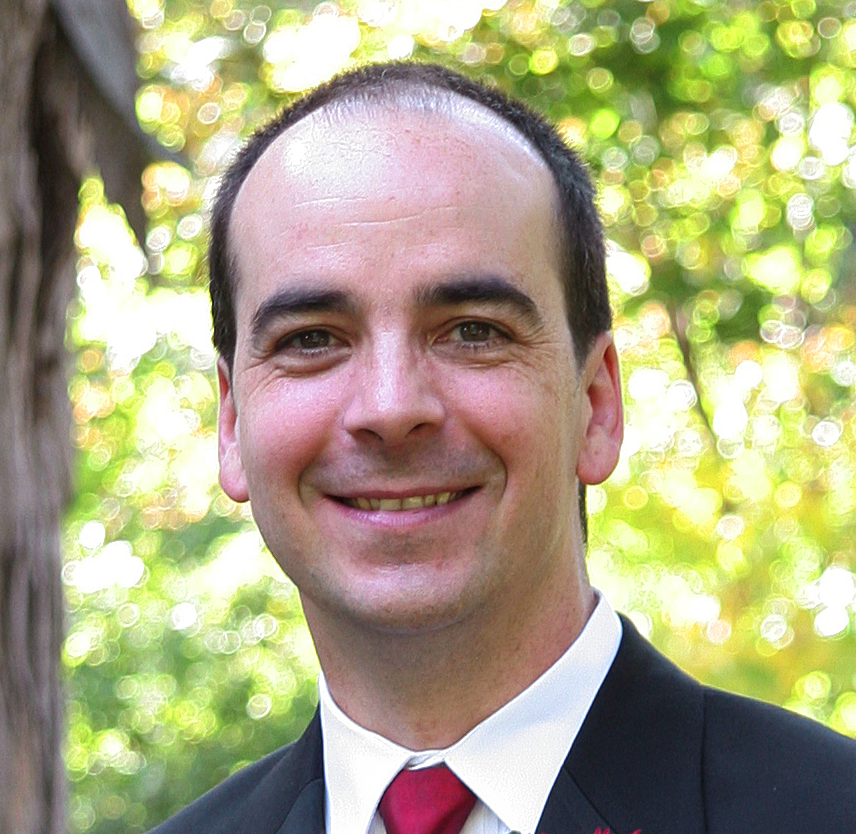}}] {Michel Bellemare} received is B.Eng. degree from Universit\'ede Sherbrooke, Sherbrooke, Qu\'ebec, Canada in Communications Engineering in 1986. He has gained experience in terrestrial and space wireless communications in various companies such as Nortel Networks, Ultra Electronics, SR Telecom, etc. He is now a Space Systems Architect at MDA corporation, Sainte-Anne-de-Bellevue, Qu\'ebec, Canada.  
\end{IEEEbiography}

\begin{IEEEbiography}[{\includegraphics[width=1in,height=1.25in,clip,keepaspectratio]{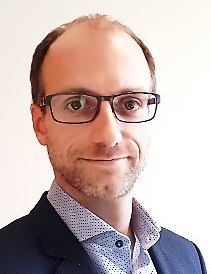}}] {Guillaume Lamontagne} obtained his B.Eng and M.Eng degrees from \'Ecole de Technologie Sup\'erieure (\'ETS), Montr\'eal, Canada in 2007 and 2009 respectively. His experience in satellite communications started through internships and research activities at the Canadian Space Agency (CSA) in 2005 and the Centre national d'\'etudes spatiales (Cnes), France, in 2006 and 2008. He joined MDA in 2009 and held various communication systems engineering and management positions before being appointed Director of Technology, Payloads in 2019. Through this role, he is leading MDA's R\&D activities for satellite communications as well as establishing the related long term development strategy.		  
\end{IEEEbiography}

\begin{IEEEbiography}[{\includegraphics[width=1in,height=1.25in,clip,keepaspectratio]{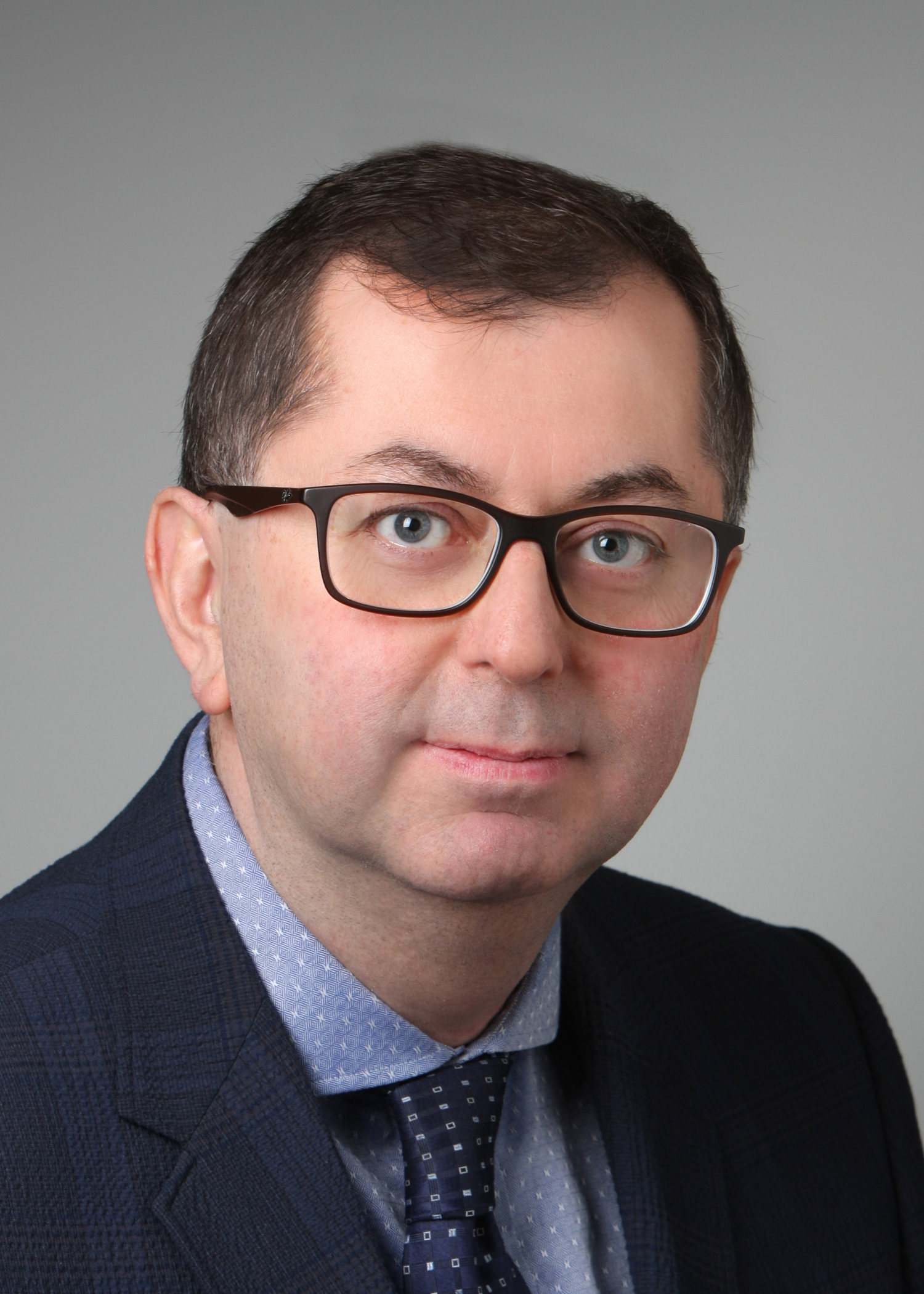}}]{Halim Yanikomeroglu}  (F'17) received the B.Sc. degree in electrical and electronics engineering from the Middle East 
	Technical University, Ankara, Turkey, in 1990, and the M.A.Sc. degree in electrical engineering and the Ph.D. degree in electrical and	computer engineering from the University of Toronto, Canada, in 1992 	and 1998, respectively. He is a Professor with the Department of Systems and Computer Engineering, Carleton University, Ottawa, Canada. His research interests cover many aspects of 5G/5G+ wireless networks. His collaborative research with industry has resulted in 37 granted 	patents. Dr. Yanikomeroglu is a fellow of the IEEE, Engineering Institute of Canada (EIC), and the Canadian Academy of Engineering (CAE). He is a Distinguished Speaker for IEEE Communications Society and IEEE Vehicular Technology Society.
\end{IEEEbiography}

\EOD

\end{document}